\documentclass[12pt,preprint]{aastex}

\usepackage{lscape}

\usepackage{times}
\usepackage{graphicx}

\shorttitle{Fermi-LAT Search for PWNe}
\shortauthors{Ackermann et al.}

\begin{document}

\title{\emph{Fermi}-LAT Search for Pulsar Wind Nebulae around gamma-ray Pulsars}


\author{
M.~Ackermann\altaffilmark{2}, 
M.~Ajello\altaffilmark{2}, 
L.~Baldini\altaffilmark{3}, 
J.~Ballet\altaffilmark{4}, 
G.~Barbiellini\altaffilmark{5,6}, 
D.~Bastieri\altaffilmark{7,8}, 
K.~Bechtol\altaffilmark{2}, 
R.~Bellazzini\altaffilmark{3}, 
B.~Berenji\altaffilmark{2}, 
E.~D.~Bloom\altaffilmark{2}
% ...
}
% ...
\altaffiltext{2}{W. W. Hansen Experimental Physics Laboratory, Kavli Institute for Particle Astrophysics and Cosmology, Department of Physics and SLAC National Accelerator Laboratory, Stanford University, Stanford, CA 94305, USA}
\altaffiltext{3}{Istituto Nazionale di Fisica Nucleare, Sezione di Pisa, I-56127 Pisa, Italy}
\altaffiltext{4}{Laboratoire AIM, CEA-IRFU/CNRS/Universit\'e Paris Diderot, Service d'Astrophysique, CEA Saclay, 91191 Gif sur Yvette, France}
\altaffiltext{5}{Istituto Nazionale di Fisica Nucleare, Sezione di Trieste, I-34127 Trieste, Italy}
\altaffiltext{6}{Dipartimento di Fisica, Universit\`a di Trieste, I-34127 Trieste, Italy}
\altaffiltext{7}{Istituto Nazionale di Fisica Nucleare, Sezione di Padova, I-35131 Padova, Italy}
\altaffiltext{8}{Dipartimento di Fisica ``G. Galilei", Universit\`a di Padova, I-35131 Padova, Italy}
% ...

\begin{abstract}
The high sensitivity of the \emph{Fermi}-LAT (Large Area Telescope) offers the first opportunity to study faint and extended GeV sources 
such as pulsar wind nebulae (PWNe). After one year of observation the LAT detected and identified three pulsar 
wind nebulae: the Crab Nebula, Vela-X and the PWN inside MSH~15$-$5\emph{2}. In the meantime, the list 
of LAT detected pulsars increased steadily. These pulsars are characterized by high energy loss rates ($\dot{E}$) 
from $\sim$$3 \times 10^{33}$ erg s$^{-1}$ to 5 $\times$ 10$^{38}$ erg s$^{-1}$ and are therefore likely to power a PWN. 
This paper summarizes the search for PWNe in the off-pulse windows of 54 LAT-detected pulsars using 16 months of survey observations. 
Ten sources show significant emission, seven of these likely being of magnetospheric origin. The detection of significant emission in 
the off-pulse interval offers new constraints on the $\gamma$-ray emitting regions in pulsar magnetospheres. 
The three other sources with significant emission are the Crab Nebula, Vela-X and a new pulsar wind nebula candidate associated with the 
LAT pulsar PSR~J1023$-$5746, coincident with the TeV source HESS J1023-575. We further explore the association between the H.E.S.S. and 
the \emph{Fermi} source by modeling its spectral energy distribution. Flux upper limits derived for the 44 remaining sources are used 
to provide new constraints on famous PWNe that have been detected at keV and/or TeV energies. 
\end{abstract}


\section{Introduction}
Since the launch of the \emph{Fermi} Gamma-Ray Space Telescope (formerly GLAST) the number of detected pulsars in the gamma-ray 
domain has dramatically increased. The list of LAT pulsars now contains 56 bright sources and certainly many more will be 
detected in the coming months. Yet most of the pulsar spin-down luminosity is not observed as pulsed
photon emission and is instead carried away as a magnetized particle wind~\citep{Gaensler2006}. The deceleration of the pulsar-driven wind 
as it sweeps up ejecta from the supernova explosion generates a termination shock at which the particles are 
pitch-angle scattered and further accelerated to ultra-relativistic energies. The pulsar wind nebula (PWN) emission, 
including synchrotron and inverse Compton components, extends across the electromagnetic spectrum from radio to TeV
energies. PWNe studies can supply information on particle
acceleration mechanisms at relativistic shocks, on the evolution of the pulsar spin-down and, at later phases, on the ambient interstellar gas. 

Despite the detection of 271 sources, EGRET could not firmly identify any PWNe besides the bright Crab Nebula. 
Most of the 170 unidentified EGRET sources at low Galactic latitudes ($|b|\leq 5 \degr$) are associated 
with star-forming regions and hence may be pulsars, PWNe, supernova remnants (SNRs), winds from massive stars, or high-mass X-ray 
binaries \citep[e.g.][]{Kaaret96,Yadigaroglu1997,Romero1999}. 
The early LAT observations~\citep{psrcat} show that \emph{Fermi} is detecting many nearby young pulsars. 
All \emph{Fermi}-LAT pulsars have a high energy loss rate ($\dot{E}$), ranging from $\sim$$3 \times 10^{33}$ erg s$^{-1}$ to 
5 $\times$ 10$^{38}$ erg s$^{-1}$. About a third of these pulsars 
are associated with pulsar wind nebulae candidates 
observed in the TeV energy range by Cherenkov telescopes. These pulsars are thus likely to power a PWN 
detectable by \emph{Fermi}. However, up to $\sim$10 GeV, the pulsed emission dominates the signal from 
the associated PWN, as can be seen with the example of Vela-X~\citep{velax}. A search for 
pulsar wind nebulae candidates around all detected \emph{Fermi}-LAT pulsars thus requires that one first removes the pulsar signal, 
thereby selecting only the unpulsed photons.

Here we report on the analysis of the off-pulse emission of 54 pulsars detected in the gamma-ray domain by 
\emph{Fermi}-LAT  using 16 months of survey observations: 45 pulsars\footnote{The pulsar PSR~J1747$-$2958 and 
its associated off-pulse emission will be studied individually due to its proximity to the Galactic center} reported in~\cite{psrcat}, the 8 
new blind search pulsars~\citep{8blind} and the millisecond pulsar PSR J0034$-$0534~\citep{j0034}. The study of the PWN in MSH~15$-$5\emph{2}, 
reported in~\cite{fermi_msh1552}, did not require the selection of off-pulse photons. Therefore, its associated pulsar PSR~B1509$-$58 is not added to our list of sources. 

The primary objective of this study is to examine the properties of the off-pulse emission of each pulsar and attempt to 
detect the potential emission associated with its pulsar wind nebula. This first population study in high energy gamma-rays allows 
us to address astrophysical questions such as:
\begin{itemize}
\item Do we see pulsar wind nebulae in all \emph{Fermi}-LAT gamma-ray pulsars ? If not, is it because of some specific 
properties of the pulsar wind or of the ambient medium ? 
\item What is the gamma-ray efficiency of PWNe and what physical parameters determine its value in addition to 
the spin-down luminosity of the pulsar ?
\item What fraction of TeV PWNe candidates are detected in the \emph{Fermi}-LAT energy range ?
\end{itemize}

The structure of the paper is as follows: section~\ref{lat} describes the LAT, sections~\ref{timing} and~\ref{ana} 
present the timing and spectral analyses, while the results are described in section~\ref{res}. 
Finally, our conclusions are summarized in section~\ref{conclu}. 

\section{LAT description and observations}
\label{lat}
The LAT is a gamma-ray telescope that detects photons by conversion into 
electron-positron pairs and operates in the energy range between 20 MeV and 300 GeV. It is made of a 
high-resolution converter tracker (direction measurement of the incident 
gamma-rays), a CsI(Tl) crystal calorimeter (energy measurement) and an 
anti-coincidence detector to identify the background of charged particles~\citep{Atwood et al. 2009}. 
In comparison to EGRET, the LAT has a larger effective area ($\sim$ 8000 cm$^{2}$ on-axis above 1~GeV), a broader 
field of view ($\sim$ 2.4 sr) and superior angular resolution ($\sim$ 
0.6$^{\circ}$ 68$\%$ containment at 1 GeV for events converting in the 
front section of the tracker). Details of the instruments and data processing 
are given in~\cite{Atwood et al. 2009}. The on-orbit calibration is described in 
\cite{VelaPulsarLAT}.

The following analysis used 16 months of data collected from 
August 4, 2008 (MJD 54682), to December 16, 2009 (MJD 55181), except for some pulsars for which portions of 
the observation period were rejected due to inadequate pulsar ephemerides, reported in Table~\ref{tab:def}. 
The {\emph{Diffuse}} class events were selected (with the tightest background rejection). From this sample, we excluded
gamma-rays with a zenith angle larger than 105$^{\circ}$ because of the possible contamination from Earth limb 
photons. We used P6$\_$V3 post-launch instrument response functions (IRFs) that take into account pile-up and accidental 
coincidence effects in the detector subsystems \footnote{See http://fermi.gsfc.nasa.gov/ssc/data/analysis/documentation/Cicerone/Cicerone\_LAT\_IRFs/IRF\_overview.html for more details}.

\section{Timing analysis}
\label{timing}
Most of the pulsars detected by \emph{Fermi}-LAT are bright point sources in the gamma-ray sky up to $\sim$10 GeV, 
though the Vela pulsar is well detected up to 25 GeV~\citep{velaII}. The study of their associated pulsar wind nebulae thus requires 
us to assign phases to the gamma-ray photons and select only those in an off-pulse window, thereby minimizing contributions from pulsars.
We phase-folded photon dates using both the \emph{Fermi} plug-in provided by the LAT team and distributed with the TEMPO2 pulsar timing package\footnote{http://sourceforge.net/projects/tempo2/}, as well as accurate timing solutions either based on radio timing observations made at the Jodrell Bank~\citep{Jodrell}, 
Nan\c cay~\citep{Nancay}, Parkes~\citep{ParkesFermiTiming} or Green Bank telescopes~\citep{GBT}, or on gamma-ray data recorded by the LAT~\citep{ray2010}. Whenever possible, data from multiple radio telescopes were combined to build timing solutions, thereby improving their accuracy and expanding their time coverage.

The origins of the timing solutions used in this analysis can be found in Table~\ref{tab:def}. For each pulsar, we list the observatories that provided the data used to build the timing model. For some pulsars, we could not produce a timing solution providing accurate knowledge of the rotational phase over the whole observation range due to glitch activity. In these cases the time intervals over which we lost phase-coherence were rejected. These intervals are given in the last column. Also listed in Table~\ref{tab:def} are the pulsar distance \citep[see][for PSR~J0248+6021]{psrcat,8blind,theureau} and the definition of the off-pulse region. These off-pulse intervals are chosen using the definition reported in previous \emph{Fermi}-LAT studies \citep{psrcat,j0034,8blind} but narrowed slightly to minimize the contamination by pulsed photons.
A few notes on these timing solutions:

\begin{itemize}
\item The rms of the timing residuals is below 0.5\% of the pulsar's rotational period in most cases, but ranges as high as 3.6\% for PSR~J1846+0919 which has one of the lowest gamma-ray fluxes. This is adequate for the analysis performed for this paper, as timing solutions are used only for rejecting pulsed photons. 

\item Glitch activity was observed for 12 pulsars over the time range considered here. These pulsars are labeled with a \textit{g} in Table~\ref{tab:def}. In all cases it was possible to model the glitch parameters in such a way that all the timing data could be used except for PSRs J0205+6449, J1413$-$6205 and J1813$-$1246 where some data had to be rejected as shown in Table~\ref{tab:def}.

\item Timing solutions were built using radio timing data for all radio-emitting pulsars except PSRs J1124$-$5916, J1741$-$2054, J1907+0602 and J2032+4127. The first is very faint in radio and was more easily timed in gamma-rays. The three others were discovered recently~\citep{camiloetal09,fermi_1907} and radio timing observations were therefore unavailable for most of the gamma-ray data considered here. For pulsars without radio emission, timing solutions were built using the data recorded by the \emph{Fermi}-LAT only.

\end{itemize}

\section{Analysis of the \emph{Fermi}-LAT data}
\label{ana}
The spectral analysis was performed using a maximum-likelihood method \citep{Mattox et al. 1996} implemented in the \emph{Fermi} 
Science Support Center science tools as the ``gtlike'' code. This tool fits a source model to the data along 
with models for the diffuse backgrounds. Owing to uncertainties in the instrument performance still under 
investigation at low energies, only events in the 100 MeV -- 100 GeV energy band are analysed. We used the map cube file gll$\_$iem$\_$v02.fit 
to model the Galactic diffuse emission together with the corresponding tabulated model isotropic$\_$iem$\_$v02.txt 
for the extragalactic diffuse and the residual instrument emission\footnote{Available from http://fermi.gsfc.nasa.gov/ssc/data/access/lat/BackgroundModels.html}. The off-pulse spectra were fit with a power-law model assuming a point-source located at the position 
of the pulsar. Nearby sources in the field of view are extracted from \cite{FirstCat} and taken into account in the study. Sources within $5^{\circ}$ of 
the pulsar of interest and showing a significant curvature index~\citep{FirstCat} were left free for the analysis assuming an exponential cut-off power-law model, while other neighbouring sources were assigned fixed power-law spectra unless the residuals showed clear indication of variability from the 1FGL catalog. 

To provide better estimates of the source spectrum and search for the best PWN candidates, we split the 
 energy range into three bands, from 100 MeV to 1 GeV, 1 to 10 GeV and 10 to 100 GeV. 
The uncertainties on the parameters were estimated using the quadratic development of the log(likelihood) around the best fit. In addition to 
the spectral index $\Gamma$, which is a free parameter in the fit, the important physical quantities are the photon flux 
$F_{0.1-100}$ (in units of ph cm$^{-2}$ s$^{-1}$) and the energy flux $G_{0.1-100}$ (in units of erg cm$^{-2}$ s$^{-1}$). 
\begin{eqnarray}
F_{0.1-100} = \int_{\rm 0.1 \, GeV}^{\rm 100 \, GeV} \frac{{\rm d} N}{{\rm d} E} {\rm d} E,\,\rm{and} \\
G_{0.1-100} = \int_{\rm 0.1 \, GeV}^{\rm 100 \, GeV} E \frac{{\rm d} N}{{\rm d} E} {\rm d} E.
\end{eqnarray}
These derived quantities are obtained from the primary fit parameters and corrected for the decreased exposure represented by the restriction to the off-pulse phase window. Their statistical uncertainties are obtained using their derivatives with respect
to the primary parameters and the covariance matrix obtained from the fitting process. 
The estimate from the sum of the three bands is on average within 30\% of the flux obtained for the global power-law fit. 

An additional difficulty with this search is that we must address cases where the source flux is not significant 
in one or all energy bands. For each off-pulse source analysed, gtlike provides the Test Statistic TS = 2$\Delta$log(likelihood) 
between models with and without the source. The TS is therefore a measure of the source significance, with TS = 25 corresponding to a significance 
of just over 4.5$\sigma$. Many sources have a TS value smaller than 25 in several bands or even in the complete energy interval. 
In such cases, we replace the flux value from the likelihood analysis by a 95\% C.L. upper limit in Tables~\ref{tab:res} and \ref{tab:res2}. 
These upper limits were obtained using the Bayesian method proposed by~\cite{helene}, assuming a photon index $\Gamma = 2$.

All fluxes and upper limits as well as the statistical uncertainties obtained using this procedure are 
summarized in Tables~\ref{tab:res} and \ref{tab:res2} and were all cross-checked 
using an analysis tool developed by the LAT team called ``Sourcelike''. In this method, likelihood fitting is iterated through the 
data set to simultaneously optimize the position and potential extension of a source, assuming spatially extended source 
models and taking into account nearby sources as well as Galactic diffuse and isotropic components in the fits. The results 
from this analysis, assuming a point-source model, are consistent with those from the likelihood analysis.

In addition to this cross-check using sourcelike, we performed a second fit to the data with gtlike incorporating the results from 
the first maximum likelihood analysis for all sources other than the one being considered, so it has a good representation of the surroundings of the source. This step returns a full Test Statistic map around each source of interest. These TS maps do not show any extended emission that could contaminate our source of interest (due to badly resolved 
diffuse background) at a TS level higher than 16.

\section{Results}
\label{res}
Pulsar wind nebulae candidates were selected using two different criteria: 
\begin{enumerate}
\item TS $>$ 25 in the whole energy range (100 MeV - 100 GeV)
\item TS $>$ 25 in one of the three energy bands (100 MeV - 1 GeV, 1 - 10 GeV, 10 - 100 GeV)
\end{enumerate}
As can be seen from Tables~\ref{tab:res} and \ref{tab:res2}, 10 of the 54 pulsars studied here satisfy one of these 
detection criteria: J0034$-$0534, J0534+2200 associated with the Crab Nebula~\citep{crab}, J0633+1746 (Geminga), J0835$-$4510 
associated with the Vela-X pulsar wind nebula~\citep{velax}, J1023$-$5746, J1813$-$1246, J1836+5925, J2021+4026, 
J2055+2539 and J2124$-$3358. A detailed study of the Crab Nebula with a model adapted to the synchrotron component 
at low energy was performed in~\cite{crab} and enabled its clear detection and identification by \emph{Fermi}-LAT. 
Similarly, a detailed morphological and spectral analysis allowed the detection of the extended emission 
from the Vela-X pulsar wind nebula~\citep{velax}. 

Aside from the Crab and Vela pulsars, J1023$-$5746 is the only candidate that shows off-pulse emission predominantly above 10 GeV, 
whereas the 7 others are mainly detected at low energy (below 10 GeV) which suggests a low energy cutoff and therefore a pulsar origin. 
To provide further details on these 7 sources and ensure that the emission detected in the off-pulse interval does not have 
a pulsar origin, we re-fitted all candidates using an exponential cutoff power-law spectral model; the results on the 
off-pulse emission of J1023$-$5746 are presented in section~\ref{j1023}. 

\subsection{Magnetospheric emission in the off-pulse window}
\label{analow}
We explored whether the exponential cutoff power-law spectral model is preferred over a simple power-law 
model by computing $TS_{\rm cutoff} = 2\Delta$log(likelihood) 
(comparable to a $\chi^2$ distribution with one degree of freedom) between the models 
with and without the cutoff. The pulsars J0633+1746, J1836+5925, J2021+4026 and J2055+2539 present a significant cutoff ($TS_{\rm cutoff} \ge 9$), J2124$-$3358 being at the edge. Pulsars with $TS_{\rm cutoff} < 9$ have poorly measured cutoff energies; in this case (for J1813$-$1246), we report in Table~\ref{tab:spec} the fit parameters assuming a simple power-law. We also determined if an 
extended uniform disk model (compared to the point-source hypothesis) better fits the data for each candidate. For this step, we used 
sourcelike and computed $TS_{\rm ext} = TS_{\rm disk} - TS_{\rm point}$.  We did not find any candidates 
with significant extension ($TS_{\rm ext} > 9$).

\begin{figure*}[ht!!]
\begin{center}
\begin{minipage}[c]{.45\linewidth}
\begin{center}
\epsscale{1.1}
\plotone{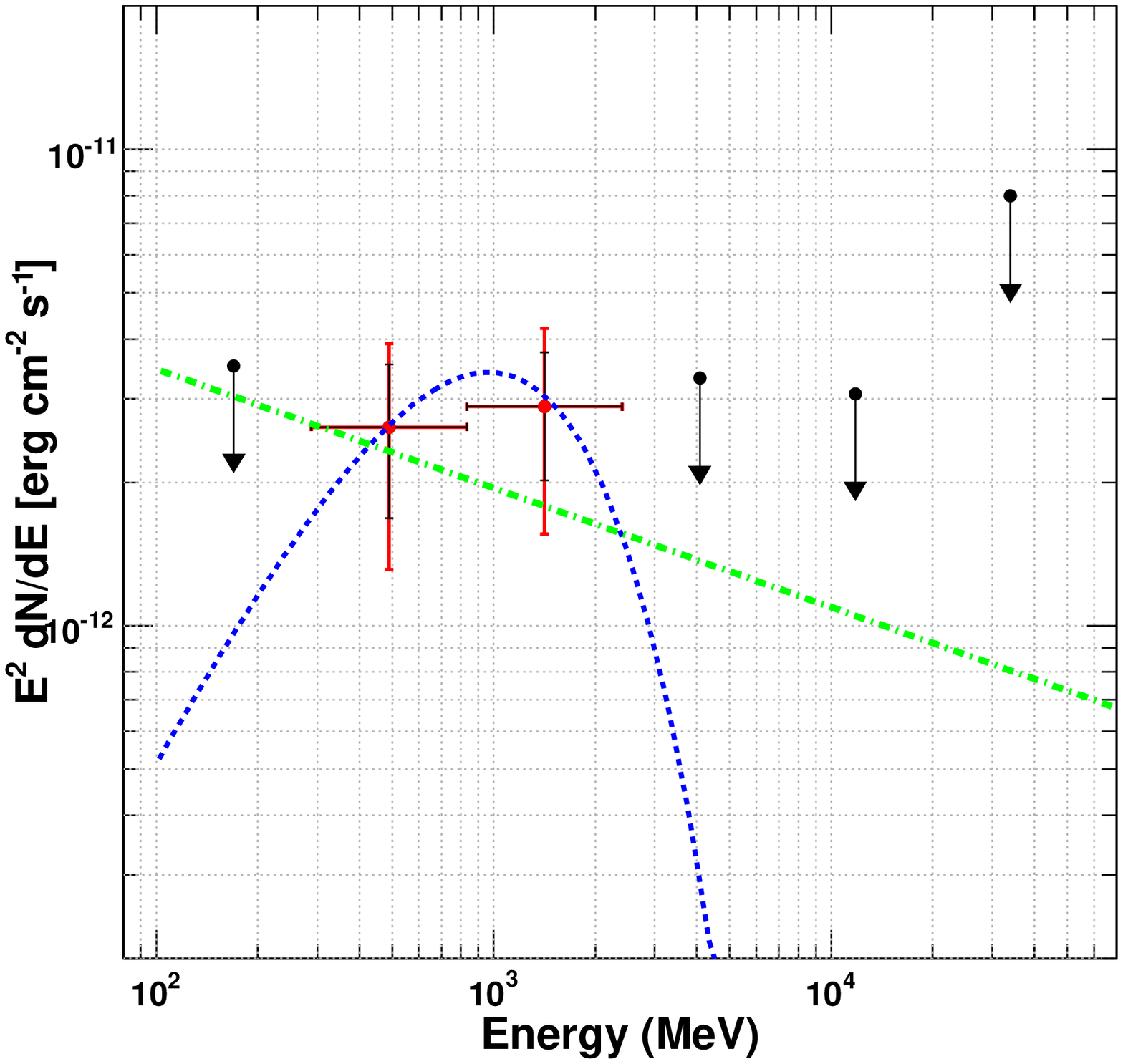}
\end{center}
\end{minipage} \hfill
\begin{minipage}[c]{.45\linewidth}
\begin{center}
\epsscale{1.1}
\plotone{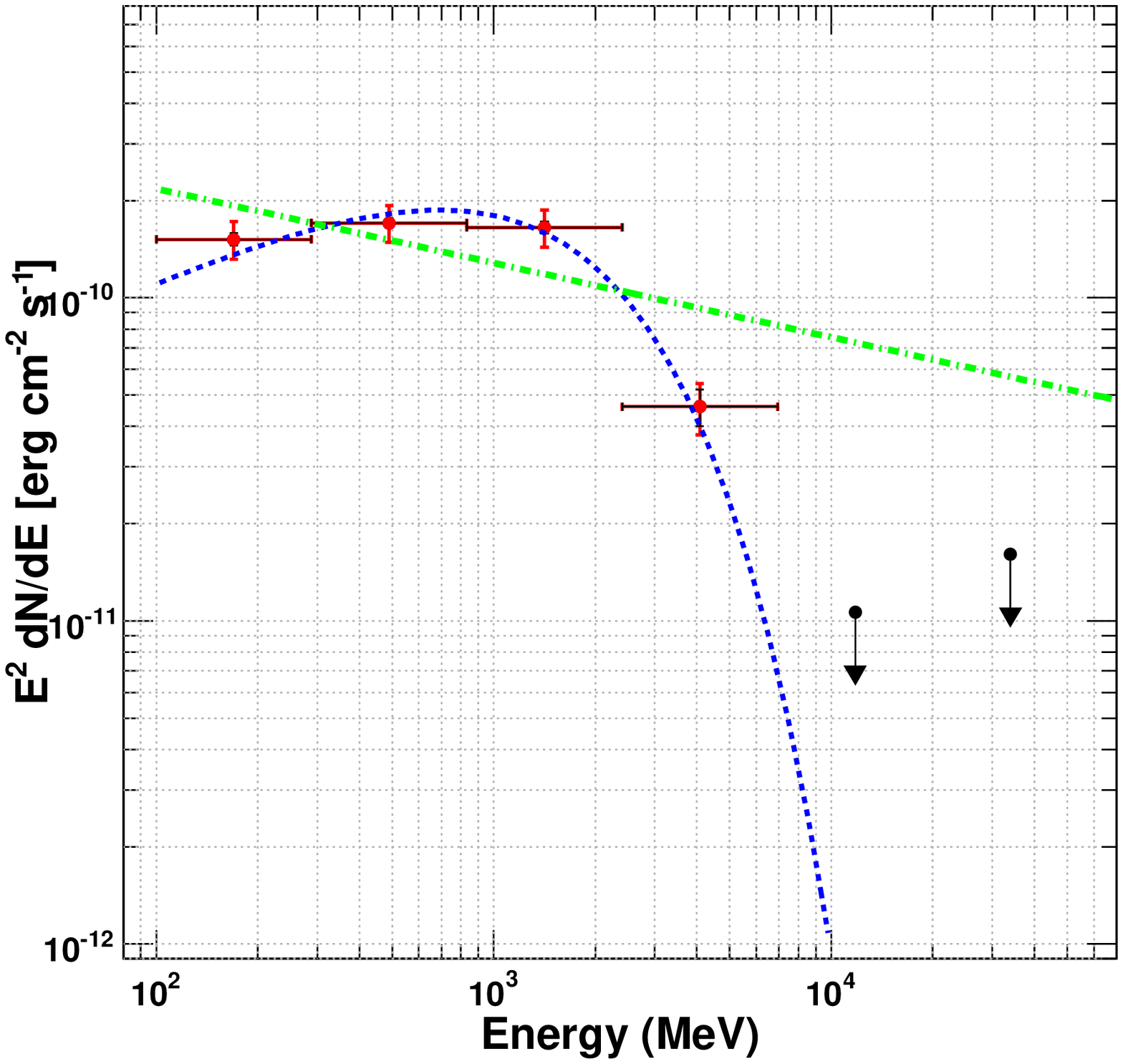}
\end{center}
\end{minipage} \vfill
\begin{minipage}[c]{.45\linewidth}
\begin{center}
\epsscale{1.1}
\plotone{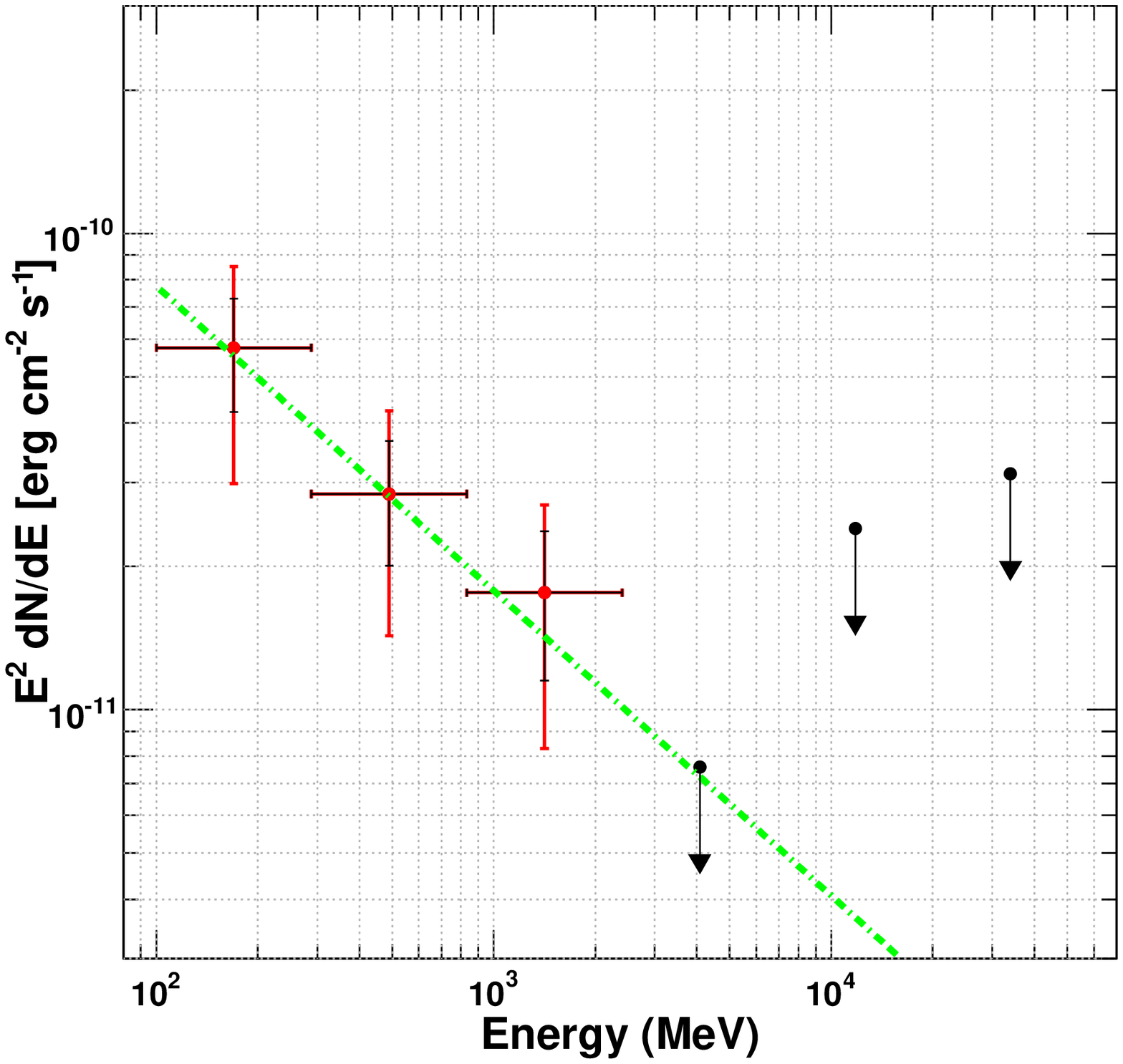}
\end{center}
\end{minipage} \hfill
\begin{minipage}[c]{.45\linewidth}
\begin{center}
\epsscale{1.1}
\plotone{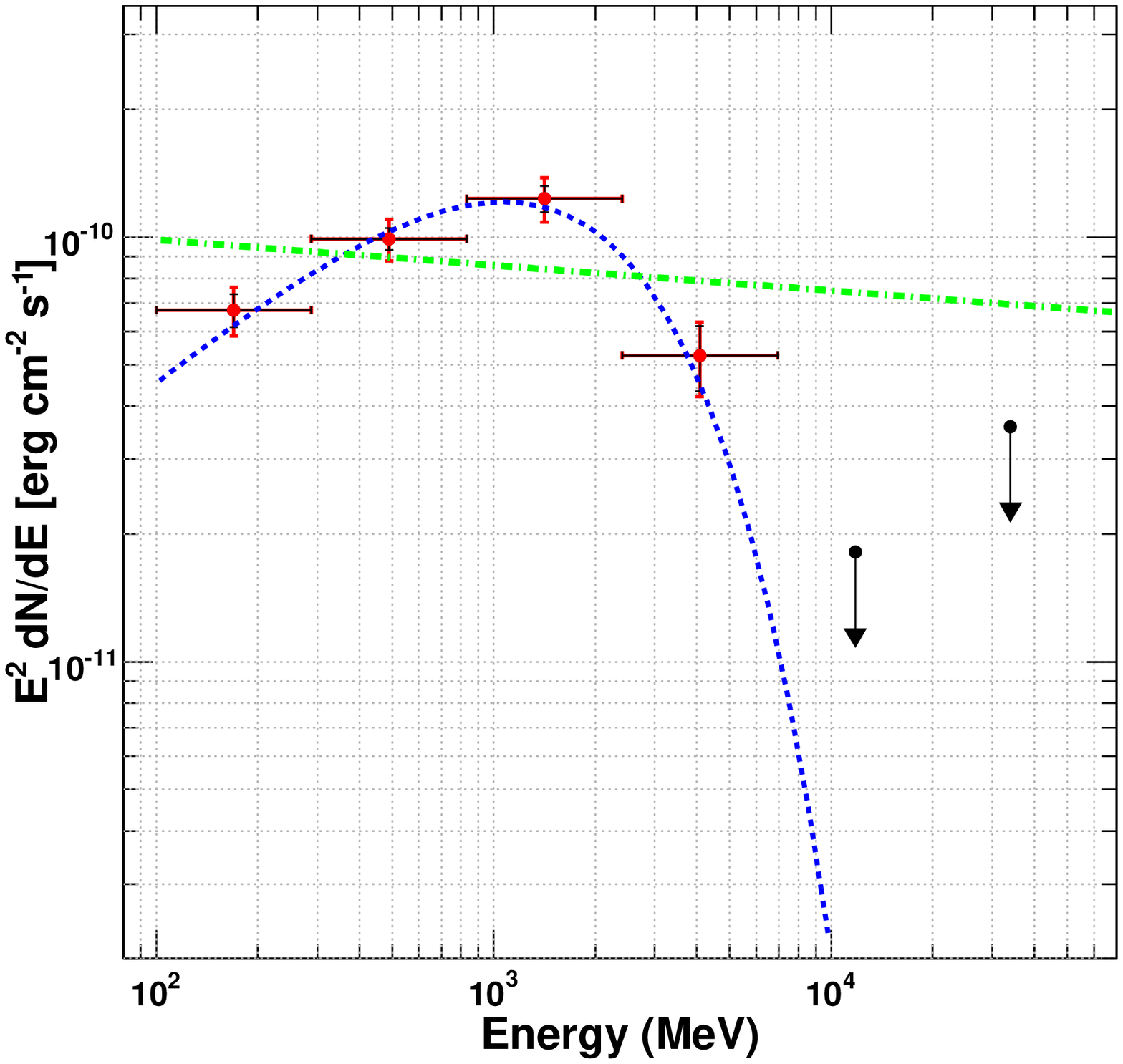}
\end{center}
\end{minipage}
\end{center}
\caption{\label{fig:sed1}Spectral energy distributions of the off-pulse emission of J0034$-$0534 (top left), J0633+1746 (top right), 
J1813$-$1246 (bottom left) and J1836+5925 (bottom right), renormalized to the total phase interval. The LAT spectral points are 
obtained using the maximum likelihood method described in section~\ref{analow} into 6 logarithmically-spaced energy bins. 
The dotted-dashed green line presents the result obtained by fitting a power-law to the data in the 100~MeV-60~GeV energy range 
using a maximum likelihood fit. The dashed blue line presents the exponential cutoff power-law model when it is favoured with respect to 
a simple power-law ($TS_{\rm cutoff}$ $\ge 9$, see section~\ref{analow}). The statistical errors are shown in black, while the red lines 
take into account both the statistical and systematic errors as discussed in section~\ref{analow}. A 95~\% C.L. upper limit is computed 
when the statistical significance is lower than 3~$\sigma$.}
\end{figure*}

\begin{figure*}[ht!!]
\begin{center}
\begin{minipage}[c]{.45\linewidth}
\begin{center}
\epsscale{1.1}
\plotone{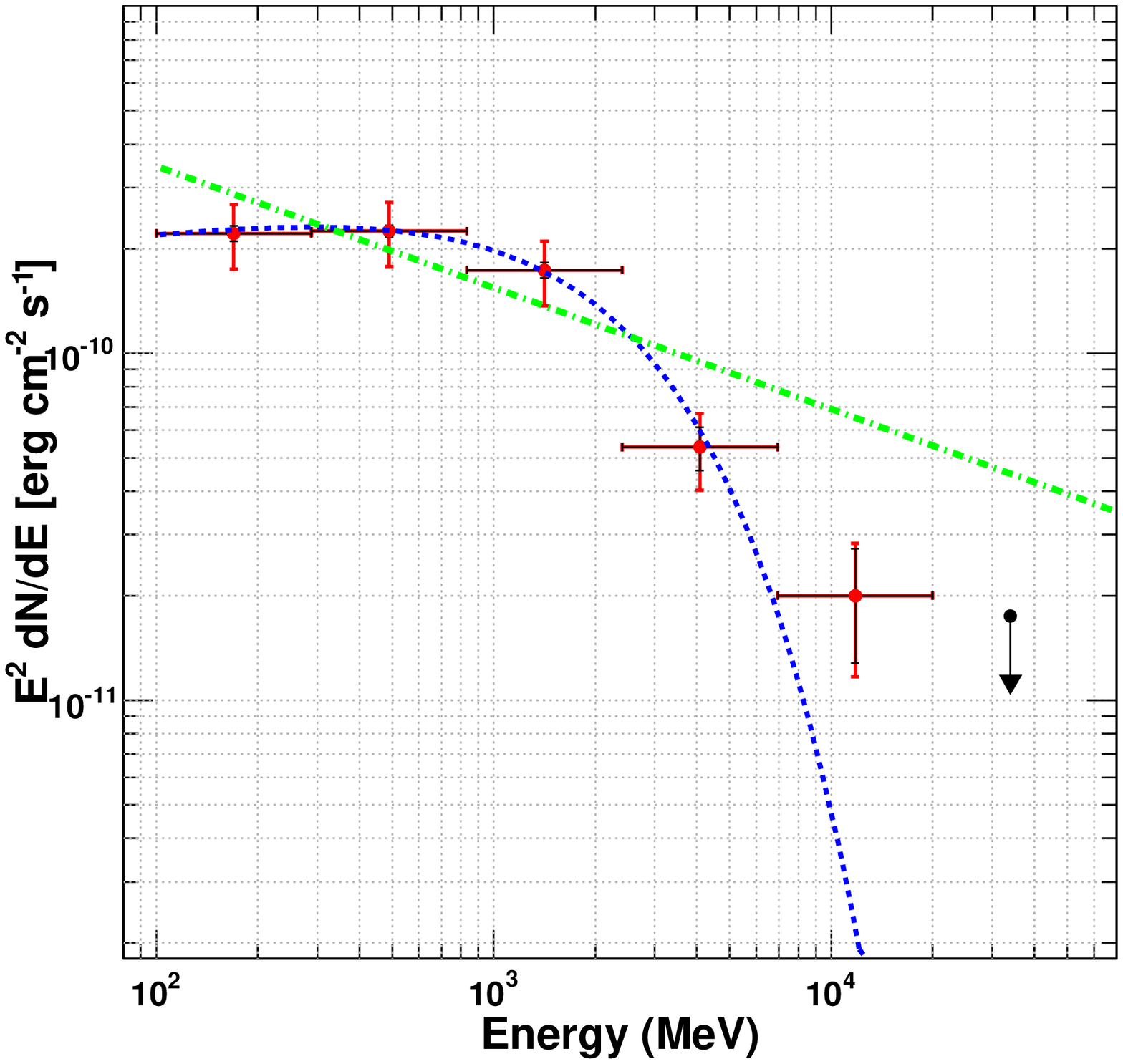}
\end{center}
\end{minipage} \hfill
\begin{minipage}[c]{.45\linewidth}
\begin{center}
\epsscale{1.1}
\plotone{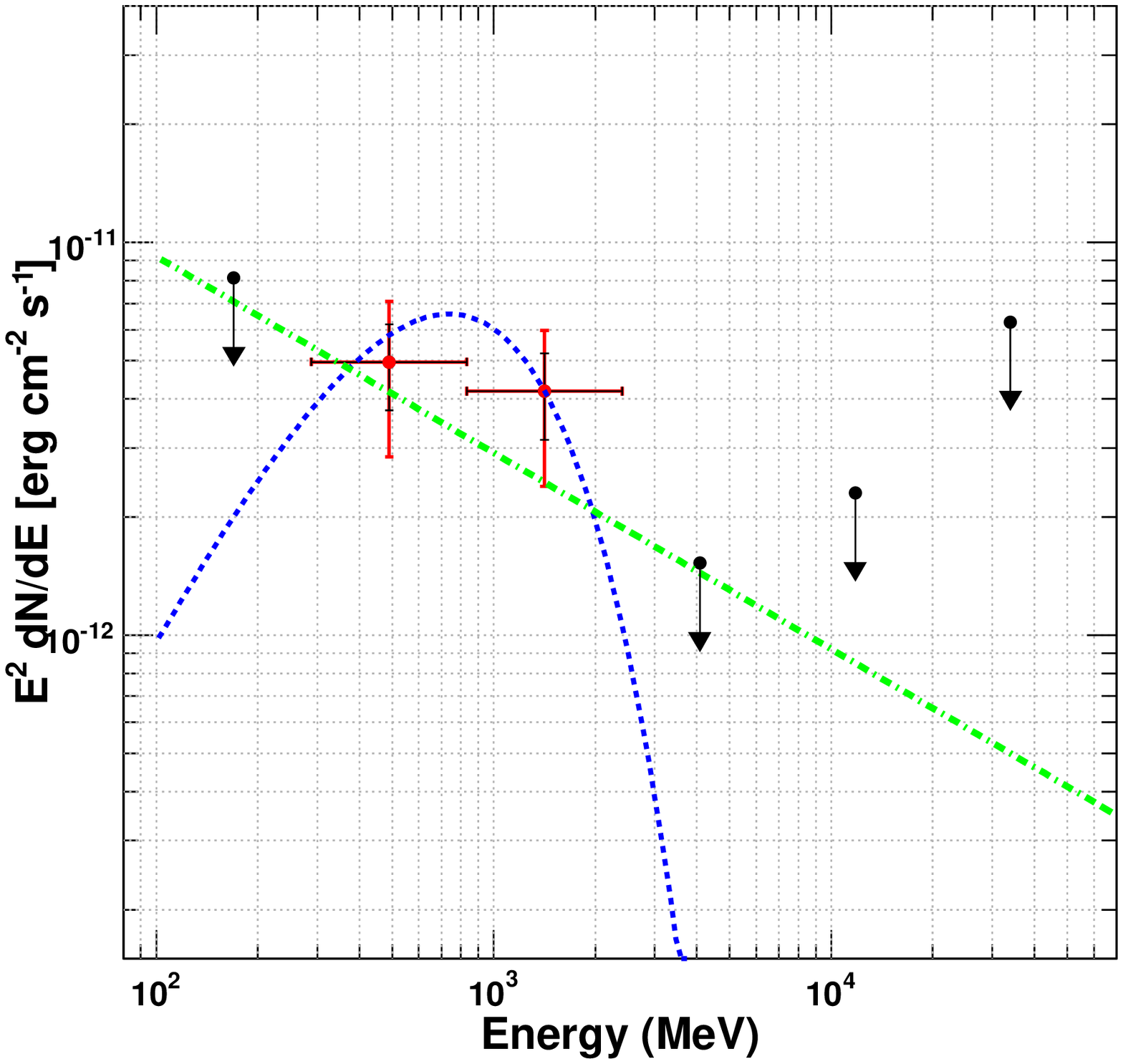}
\end{center}
\end{minipage} \hfill
\begin{minipage}[c]{.45\linewidth}
\begin{center}
\epsscale{1.1}
\plotone{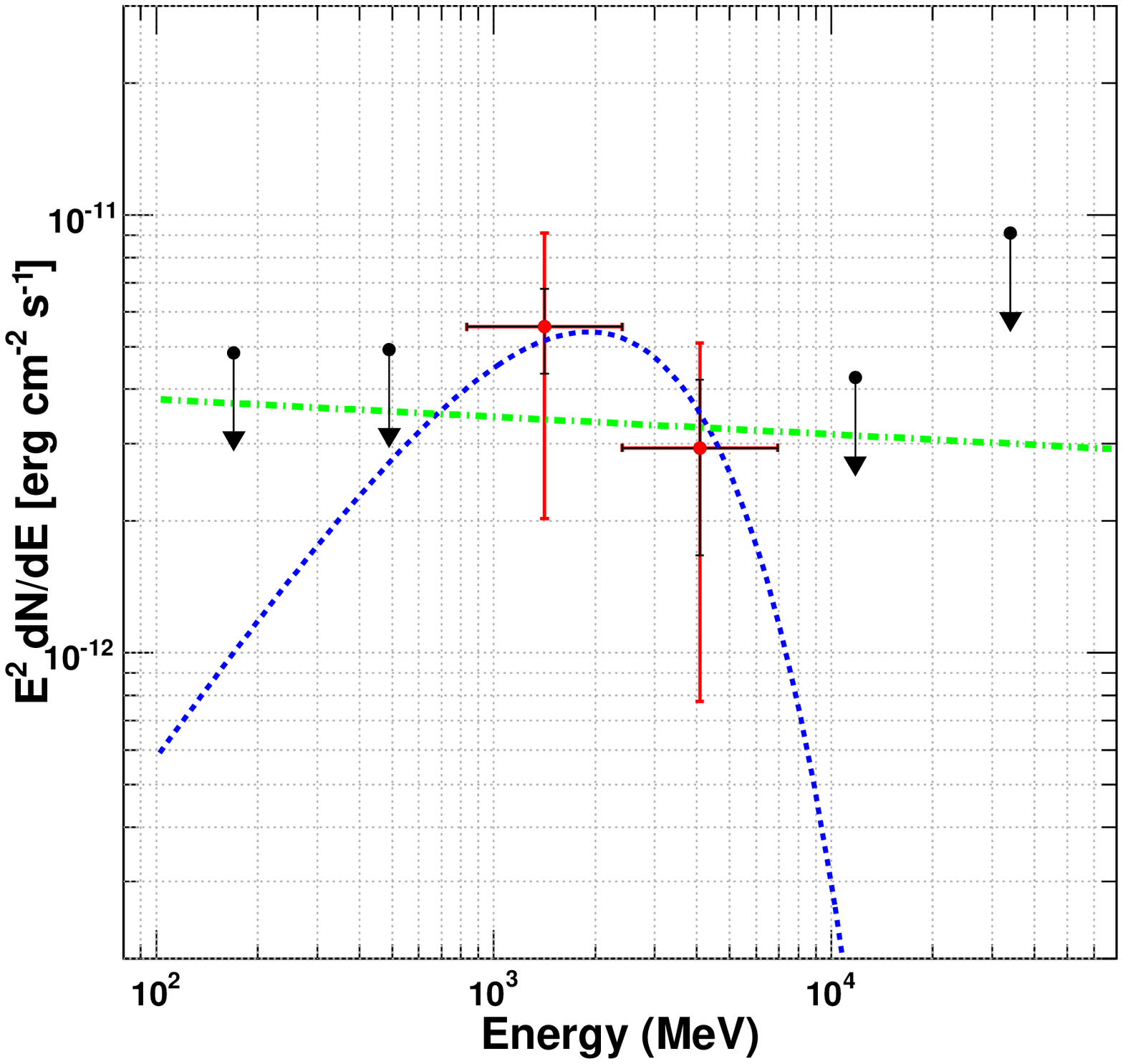}
\end{center}
\end{minipage}
\end{center}
\caption{\label{fig:sed2}Spectral energy distributions of the off-pulse emission of J2021+4026 (top left), 
J2055+2539 (top right) and J2124$-$3358 (bottom), renormalized to the total phase interval. 
Same conventions as for Figure~\ref{fig:sed1}.}
\end{figure*}

The \emph{Fermi}-LAT spectral points for each source listed in Table~\ref{tab:spec} were obtained by dividing the 100 MeV -- 60 GeV 
range into 6 logarithmically-spaced energy bins and performing a maximum likelihood spectral 
analysis in each interval, assuming a power-law shape for the source with a fixed photon index. The results, renormalized to the total phase interval, are 
presented in Figures~\ref{fig:sed1} and~\ref{fig:sed2} together with the maximum likelihood fit in the whole energy range, 
assuming an exponential cutoff power-law (dashed blue line) or a power-law (dotted-dashed green line). This analysis is more reliable than a 
direct fit to the spectral points of Figure~\ref{fig:sed1} and \ref{fig:sed2} since it accounts for Poisson statistics of the data.

Three different systematic uncertainties can affect the results derived with this analysis. The main systematic 
at low energy is due to the uncertainty in the Galactic diffuse emission. Different versions of the Galactic 
diffuse emission generated by GALPROP were used to estimate this error in the case of the supernova remnants W51C and W49 
\citep{W51C,W49B}. The difference with the best fit diffuse model is
found to be $\le 6$\%. Therefore, we estimated this systematic error by changing the normalization of the Galactic diffuse model 
artificially by $\pm 6$\%. The second uncertainty, common to every source analyzed with the LAT, is due to the uncertainties in the effective area. 
This systematic is estimated by using modified instrument response functions 
(IRFs) whose effective area bracket that of our nominal IRF. These `biased' IRFs 
are defined by envelopes above and below the nominal dependence of the effective 
area with energy by linearly connecting differences of (10\%, 5\%, 20\%) at 
log(E) equal to (2, 2.75, 4), respectively. The third systematic is related to the morphology and spectrum of the source. 
Taking a power-law spectral shape and a point-source morphology at the pulsar position are 
strong assumptions that can affect the flux and the spectral indices of the off-pulse component derived with this simple analysis, as has 
been demonstrated for the case of the Vela-X pulsar~\citep{velax}. A more detailed analysis of each source is beyond 
the scope of this paper and must be handled on a case by case basis.
We combine the other two systematic errors in quadrature to estimate the total systematic error at each energy and propagate it through 
to the fit model parameters reported in Table~\ref{tab:spec}.

The lack of extended emission and the significant spectral cutoffs at low energies (from 0.43 to 1.71 GeV) suggest that the 
off-pulse emission detected by \emph{Fermi}-LAT is lkely magnetospheric and that we do not observe pulsar wind nebulae 
for J0633+1736, J1836+5925, J2021+4026, J2055+2539 and J2124$-$3358. 
This was already suggested in previous \emph{Fermi}-LAT publications on the first two pulsars, J0633+1746~\citep{geminga} 
and J1836+5925~\citep{j1836}.

The cases of the \emph{Fermi}-LAT pulsar PSR~J1813$-$1246 and the millisecond pulsar J0034$-$0534 are harder to handle due to the limited statistics. For J0034$-$0534 an unpulsed component of emission 
from particle acceleration in the wind termination shock might be expected since this pulsar is in a binary system,
though the two-pole caustic model also predicts a faint 
signal in the off-pulse window of J0034$-$0534. In the case of J1813-1246, which shows a steep spectrum with no significant 
cutoff, we cannot rule out the PWN origin with the current statistics. We therefore cannot definitely determine the origin of the emission detected by \emph{Fermi}-LAT for these two candidates.

\subsection{A plausible pulsar wind nebula candidate powered by PSR~J1023$-$5746}
\label{j1023}
\subsubsection{\emph{Fermi}-LAT results on the off-pulse emission of PSR~J1023$-$5746}
\label{res1023}
In 2007, H.E.S.S. reported the detection of very high-energy gamma-rays from an extended source, HESS J1023$-$575, in the direction of the young stellar 
cluster Westerlund 2~\citep{westerlund2}. Four scenarios to explain the TeV emission were suggested: colliding stellar winds in the 
WR 20a binary system (although this scenario can hardly reproduce the observed source extension of 0.18$^{\circ}$), collective effects of stellar winds 
in the Westerlund 2 cluster (although the cluster angular extent is smaller than that of the very high energy gamma-ray emission), diffusive shock 
acceleration in the wind-blown bubble itself, and supersonic winds breaking out into the interstellar medium. 
Recently, \emph{Fermi}-LAT discovered the very young (characteristic age of 4.6~kyr) and energetic (spin-down power of $1.1 \times 10^{37}$~erg~s$^{-1}$) 
pulsar J1023-5746, coincident with the TeV source HESS J1023$-$575~\citep{8blind}. 

As noted above, J1023$-$5746 is the only candidate that does not show any 
off-pulse emission below 10 GeV, whereas its signal above 10 GeV is $> 3\sigma$. Therefore, an exponential cutoff power-law 
model, as used for the 7 other candidates, will not represent the data properly. For these reasons, we decided to analyze 
this source separately.

We searched for a significant source extension using sourcelike with a uniform disk hypothesis (compared to the point-source hypothesis). 
The difference in TS between the uniform disk and the point-source hypothesis is negligible 
which demonstrates that the two models fit equally well with the current limited statistics. We have also examined the correspondence of 
the gamma-ray emission with different source shapes by using gtlike with assumed multi-frequency templates. 
For this exercise we compared the TS values of the point source, uniform disk and Gaussian spatial models with
values derived when using a morphological template from the H.E.S.S. gamma-ray excess map~\citep{westerlund2}. We did not find 
any significant improvement (difference in TS $\sim$ 3) between the different models and we therefore cannot rule out a simple point 
source morphology.

\begin{figure*}[ht!!]
\begin{center}
\end{center}
\epsscale{0.9}
\plotone{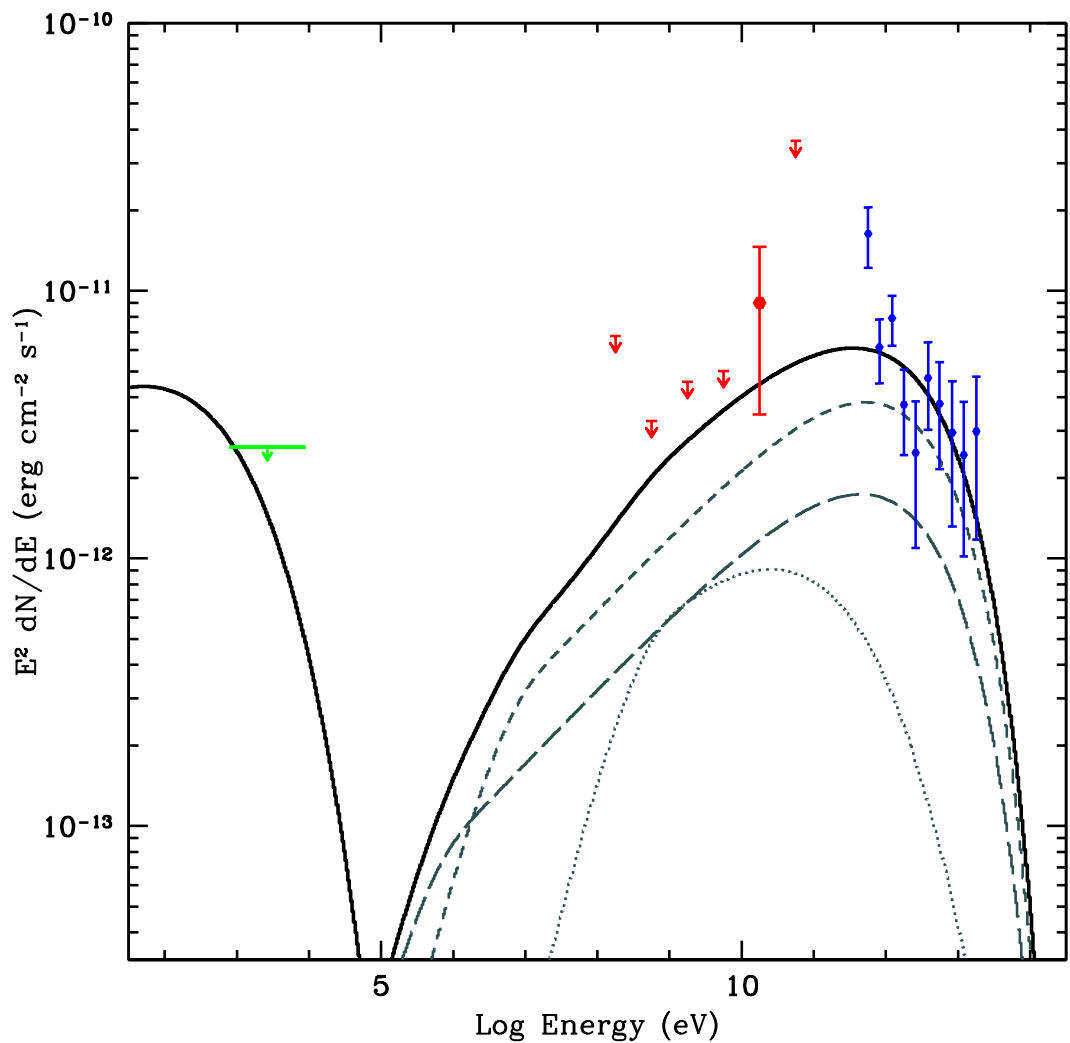}
\caption{\label{fig:sed1023model}Spectral energy distributions of the off-pulse emission of PSR~J1023$-$5746. 
The LAT spectral points (red) are obtained using the maximum likelihood 
method described in section~\ref{res1023} in 7 logarithmically-spaced energy bins. A 95~\% C.L. upper limit is computed when the statistical significance is lower than 3~$\sigma$. The blue points represent the H.E.S.S. spectral points~\citep{westerlund2}. 
The Suzaku upper limit is shown with a green arrow \citep{fujitaetal09}. 
The black line denotes the total synchrotron and Compton emission from the nebula as described in 
section~\ref{mod1023}. Thin curves indicate the Compton components from scattering on the CMB (long-dashed), IR (medium-dashed), and stellar (dotted) photons.}
\end{figure*}

To further investigate the
off-pulse spectrum and avoid reliance on a given spectral shape, we derived the spectral 
points by dividing the 100 MeV -- 100 GeV range into 6 logarithmically-spaced energy bins and performing a maximum likelihood spectral 
analysis in each interval assuming a point source at the position of the pulsar (as explained in section~\ref{analow}). The result, 
renormalized to the total phase interval, is presented in Figure~\ref{fig:sed1023model} with a red point and arrows. The signal is only significant above 10 GeV 
and is consistent with the H.E.S.S. spectral points. 

\subsubsection{Broad-band modeling}
\label{mod1023}
The connection between the GeV flux as observed by \emph{Fermi} and the TeV flux as seen by H.E.S.S. supports a common origin for the gamma-ray 
emission. The extension of the H.E.S.S. source, the off-pulse \emph{Fermi} signal, and the energetics of this young pulsar point towards a pulsar wind nebula origin. The very large number of PWNe detected in the TeV energy range 
(the most numerous class of Galactic TeV sources) and the significant number of PWNe associated with \emph{Fermi}-LAT pulsars make this 
scenario highly probable.\\
Analysis of CO emission and 21 cm absorption along the line of sight to 
Westerlund 2 gives a kinematic distance of $6.0 \pm 1.0$ kpc to the star cluster \citep{dame07}.  
The assumption that TeV emission stems from the pulsar associating PSR~J1023$-$5746 with
Westerlund 2 is problematic, however. The $8 \arcmin$ separation of the pulsar and the cluster imply 
an extremely high transverse velocity of $\sim 3000 \, \rm{km \, s^{-1}}$ for a 6 kpc distance and the pulsar's
characteristic age of 4.6 kyr.  In addition, the 
 $0.18^{\circ}$ extension of HESS J1023$-$757 is equivalent to $19$ pc at a distance of 6~kpc, 
which predicts a very fast mean expansion velocity of 
$4000 \, {\rm km \, s^{-1}}$ over 4.6 kyr. 
The pulsar pseudo-luminosity distance places it much closer at 2.4 kpc (based on inferred beaming and gamma-ray efficiencies),
though the scatter in inferred luminosities in radio-loud LAT pulsars translates to uncertainties in this estimate
of the order of factors of 2-3 \citep{8blind}. Both pulsar efficiency and PWN expansion velocity would be 
anomalously high at 6 kpc, so we adopt the pseudo-distance of 2.4 kpc.  
At this distance the pulsar spin-down power ($1.1\times10^{37} \, {\rm erg \, s^{-1}}$) can
easily account for the VHE luminosity above 380 GeV of $1.4\times10^{34} \, d_{2.4}^{2} \, {\rm erg \, s^{-1}}$. 

At longer wavelengths the vicinity of Westerlund 2 has undergone extensive study.
Archival Chandra data indicate a faint source coincident with PSR J1023$-$5746, with an X-ray index of 
$\Gamma = 1.2 \pm 0.1$ and unabsorbed $0.5-8$ keV flux of 
$1.3_{-0.3}^{+0.5} \times 10^{-13} \, \rm{erg \, cm^{-2} \, s^{-1}}$, though this does not affect modelling
of the extended nebula.
Recent Suzaku observations \citep{fujitaetal09} found no sign of diffuse non-thermal emission within the TeV contours,
and placed a $0.7-2$ keV upper limit on the diffuse flux from the entire XIS field of view of 
$2.6 \times 10^{-12} \, {\rm erg \, cm^{-2} \, s^{-1}}$. Fujita and collaborators also note that it is unlikely that 
strong X-ray emission extends beyong this field since their upper limit is consistent with the one derived using the wide HXD field (34' $\times$ 34'). 
Investigations of molecular clouds toward Westerlund 2 \citep{fukuietal09} show features of a few 
$\times 10^4 M_{\odot}$, though
CO observations indicate a low density of gas (likely $n < 1 \, {\rm cm^{-3}}$) in the region that coincides 
with the bulk of the TeV emission. 
Radio observations of RCW 49 (the H~{\sc II} complex surrounding
Westerlund 2) found a flux of 210 Jy at 843 MHz in the core \citep{whiteoak+uchida97}; this provides a non-constraining
upper limit on the radio flux corresponding to the gamma-ray source.  

We computed SEDs (Spectral Energy Distributions) from evolving electron populations over the lifetime of the pulsar in a series
of time steps, as described in \citep{velax}. 
As pulsars spin down, they dissipate rotational kinetic energy via 
$$
\dot{E} = I \Omega \dot{\Omega}
\eqno (1)
$$ 
with $\Omega$ the angular frequency and $I$ the neutron star's moment of inertia, assumed to be $10^{45} \, \rm{g \, cm^2}$.  
This energy goes into a magnetized particle wind, and for magnetic dipole spindown of the pulsar
$$
\dot{\Omega} \propto \Omega^3 
\eqno (2)
$$
Integrating equation 2 yields the age of the system \citep{1977puls.book}: 
$$
T = \frac{P}{2\dot{P}} \left( 1- \left( \frac{P_0}{P} \right)^2 \right)
\eqno (3)
$$
where $P_0$ is the initial spin period, $\dot{P}$ the period derivative.
For $P_0 \ll P$ this equation reduces to the charactistic age of the pulsar $\tau_c \equiv P/2\dot{P}$.
The spin-down luminosity of the pulsar evolves as \citep{pc73}:
$$ 
\dot{E} = \dot{E_0} (1 + \frac{t}{\tau_{0}})^{-2}
\eqno (4)
$$
with the initial spin-down timescale defined as
$$
\tau_{0} \equiv \frac{P_0}{2 \dot{P_0}}
\eqno (5)
$$ 
with $\dot{P_0}$ the initial spin period derivative. 
Given that the current $P$, $\dot{P}$, and $\dot{E}$ are known, once an initial period is selected 
the age and spin-down history of the system is determined according to the equations above. 
We assume a particle dominated wind 
such that the wind magnetization parameter $\sigma \sim 10^{-3}$. 
Therefore the power injected in the form of electron/positron pairs is $\dot{E_e} = 0.999 \dot{E}$.

As the distribution of particles expand with the PWN, they lose energy through adiabatic cooling, though synchrotron cooling
typically dominates for the earliest phase of PWNe evolution. 
We assume that the 
radius $R$ of the PWN scales linearly with time, and we select a magnetic field dependence of $B \propto t^{-1.5}$.
Both these behaviors closely mimic the behavior of $B$ and $R$ computed by 
\citet{gelfandetal09} for early stage PWN evolution 
prior to the compression and reexpansion phases caused by the interaction of the reverse shock.
Selection of appropriate photon fields is crucial to accurate determination of IC fluxes.  
We therefore follow \cite{pms06} in
estimating photon fields (CMBR, dust IR, and starlight) 
at the appropriate Galactic radii, unless local studies provide better estimates than these
Galactic averages. 

To compute the PWN SED we inject at each time step a power-law spectrum of relativistic
electrons with a high energy exponential cutoff. 
We also employ a low energy cutoff of 10~GeV for the electron spectrum,
which is within the realm of minimum particle energies considered by \citet{kc84}.
The energy content of this particle population varies with time following the pulsar spin down (eq. 4), 
though we treat the index and cutoff energies as static. We then adjust the size and magnetic field according
to the models described above. Finally, we calculate the subsequent particle spectrum at time $t+\delta t$
by calculating the energy loss of the particles due to  
adiabatic losses as well as radiation losses from synchrotron and IC (including Klein-Nishina effects). 
Injection (and evolution) occurs in time steps much smaller than the assumed age.

Model fitting is achieved by minimizing the $\chi^{2}$ between model and data
using the downhill simplex method described in \citet{pressetal92}.
We consider three variables: the initial spin period, electron slope, and high energy electron cutoff.
With only an X-ray upper limit, the mean magnetic field within the gamma-ray source is poorly constrained, so we fix the
current magnetic field to $5 \, \mu \rm{G}$ (which is the best value obtained when we allow the magnetic field to vary), or  $\sim 2$ mG at pulsar birth. 
For each ensemble of these 3 variables we evolve the system over the pulsar 
lifetime and calculate $\chi^{2}$.  The simplex routine subsequently varies the parameters of interest
to minizimize the fit statistic.  
We estimate parameter errors by computing $\chi^{2}$ for a sampling of 
points near the best fit values and using these points to fit the $3-$dimensional ellipsoid 
describing the surface of $\Delta \chi^{2} = 2.71$.  Under the assumption of Gaussian errors, the minima and maxima of this surface give the 90\% 
errors of the parameters. 

For the assumed Galactic radius of PSR J1023$-$5746, dust IR photons typically peak at 
$\approx T = 30$ K with a density $\approx 1$ eV$\rm \, cm^{-3}$, 
while stellar photons peak at $\approx T = 2500$ K with a density $\approx 2$ eV$\rm \, cm^{-3}$ 
\citep{pms06}. 
With these photon fields (and CMBR) we apply the model described above.
Figure~\ref{fig:sed1023model} indicates that 
IR photons dominate IC scattering above 10 GeV, with all three photon fields contributing for lower energies. 
For the best fit we find $\chi^2=13.7$ for 8 degrees of freedom, with an electron power law index of $2.44 \pm 0.06$, 
high energy cutoff at $60 \pm 45$~TeV, and initial spin period of $63 \pm 17$~ms. 
These parameters imply $\approx3\times10^{48}$ erg have been injected in the form of electrons, and an age of 3100 years.  

A hadronic origin for the observed gamma rays is also possible, and we follow \citet{ket06} in calculating 
the photons from proton-proton interactions and 
subsequent $\pi^0$ and $\eta$-meson decay. Proton-proton interactions also yield $\pi^{\pm}$ mesons 
which decay into secondary electrons, which we evolve in time. 
The timescale for pion production via p-p interactions is given by
$ \tau_{pp} \approx 1.5 \times 10^{8} \, (n/1 \, \rm cm^{-3})^{-1}$ years 
\citep{b70};
this timescale is significantly greater than the expected age of the system, so the proton spectrum 
is treated as static. We are able to fit the gamma-ray data only if the energy in protons exceeds
$2 \times 10^{50} \, (n/1 \, \rm cm^{-3}) \, d_{2.4}$ erg, with a $\chi^2 \approx 15$ for 8 degrees of freedom.  
A hadronic origin for the gamma-rays is therefore energetically disfavored unless the gas
density is much greater than $1 \, {\rm cm^{-3}}$ throughout the bulk of the VHE emitting region. 
Yet we cannot rule out such an origin in the confused region around Westerlund 2, even though a 
PWN origin is reasonable given the fit parameters discussed above. 

Independent of the origin of the gamma rays, the lack of X-rays from the immediate 
vicinity of PSR J1023$-$5746 is perplexing given its extremely high spin-down luminosity. 
One possibility 
is that electrons rapidly escape from
the inner nebula into a low pressure bubble with correspondingly low magnetic field.
For an electron conversion efficiency of $\sim 1$,
at the current $\dot{E}$ after a mere $\approx 2$ years enough electrons are present 
in the inner nebula
to recreate the observed X-ray flux for a $20 \, {\mu {\rm G}}$ field appropriate
for a termination shock. 
This timescale is comparable to the time for particles to reach the termination shock.
Post-shock flow in PWNe, as determined by torus fitting, is typically $\approx 0.7 \, c$ \citep{ng+romani08};  
at this velocity particles will traverse the $\sim 8 \arcsec$ X-ray nebula surrounding J1023 in $\sim 0.5 d_{2.4}$ year. 


\section{Discussion}
\label{conclu}
\subsection{Constraints on pulsar modeling}
The high-quality statistics obtained with the \emph{Fermi}-LAT both on the light curves and the spectra of the 54 pulsars detected
allow a more detailed comparison with theoretical models than previously possible. The detection or lack of significant emission in 
the off-pulse interval can also be used to discriminate between the different models. Currently, there are two classes of models that 
differ in the location of the emission region. The first comprises polar cap (PC) models which place the emission near the magnetic 
poles of the neutron star \citep{Daugherty and Harding 1996}. 
The second class of outer magnetosphere models consists of the outer 
gap (OG) models \citep{Romani 1996}, in which the emission extends between the null charge surface and the light cylinder, 
the two-pole caustic (TPC) models \citep{Dyks and Rudak 2003} which might be realized in slot gap (SG) acceleration models 
\citep{MuslimovHarding2004}, in which the emission takes place between the neutron star surface and the light cylinder 
along the last open field lines, separatrix layer (SL) models \citep{Bai09}, in which emission
takes place from the neutron star surface to outside the light cylinder and finally pair-starved polar cap (PSPC) 
models \citep{MuslimovHarding2004}, where emission takes place throughout the entire open field region. 
Observations by \emph{Fermi} of simple exponential cutoffs in the spectrum of Vela and other bright pulsars \citep{VelaPulsarLAT,velaII}, 
instead of super-exponential cutoffs expected in PC models, have clearly ruled out this class of model for \emph{Fermi} pulsar emission. 
The outer magnetosphere models make different predictions for the level of off-pulse emission. Classic OG models 
\citep{RomYad1995,Cheng2000,Romani2010}, for which there is no emission 
below the null charge surface, predict no off-pulse emission except at very small inclination angles and large viewing angles near 90 degrees. 
TPC models predict pulsed emission over most of the rotational phase at a level that depends on inclination, viewing angle, and gap width \citep{Venter2009,Romani2010}.
In general, light curves for larger gap widths, expected for middle-aged and older pulsars in the SG model and when the viewing direction 
makes a large angle to the magnetic axis, have higher levels of off-pulse emission. 
The force-free magnetosphere SL model~\citep{Bai09} also predicts light curves with off-pulse emission, since some radiation 
in this case also comes from below the null surface. PSPC models are expected to operate in old and millisecond pulsars and 
predict off-pulse emission as well \citep{Venter2010}.

Among the 54 pulsars analyzed in this paper, only 10 show a significant signal in their off-pulse, 7 of which are likely of 
magnetospheric origin (J0034$-$0534, J0633+1746, J1813$-$1246, J1836+5925, J2021+4026, J2055+2539, J2124$-$3358). 
Two of the 7 showing off-pulse emission, J0034$-$0534 and J2124$-$3358, are millisecond pulsars. J1836+5925 with 
$\dot E = 1.2 \times 10^{34}\rm \, erg\,s^{-1}$, J2055+2539 with $\dot E = 5 \times 10^{33}\rm \, erg\,s^{-1}$ 
and J0633+1746 (Geminga) with $\dot E = 3.3 \times 10^{34}\rm \, erg\,s^{-1}$ have among the lowest spin-down 
luminosities of the normal Fermi detected pulsars. While J1813$-$1246 and J2021+4026 have higher $\dot E$ ($6.3 \times 10^{36}\rm \, erg\,s^{-1}$ and $1.1 \times 10^{35}\rm \, erg\,s^{-1}$ respectively), both have unusually wide gamma-ray pulses.

\begin{figure*}[ht!!]
\begin{center}
\begin{minipage}[c]{.45\linewidth}
\begin{center}
\epsscale{1.1}
\plotone{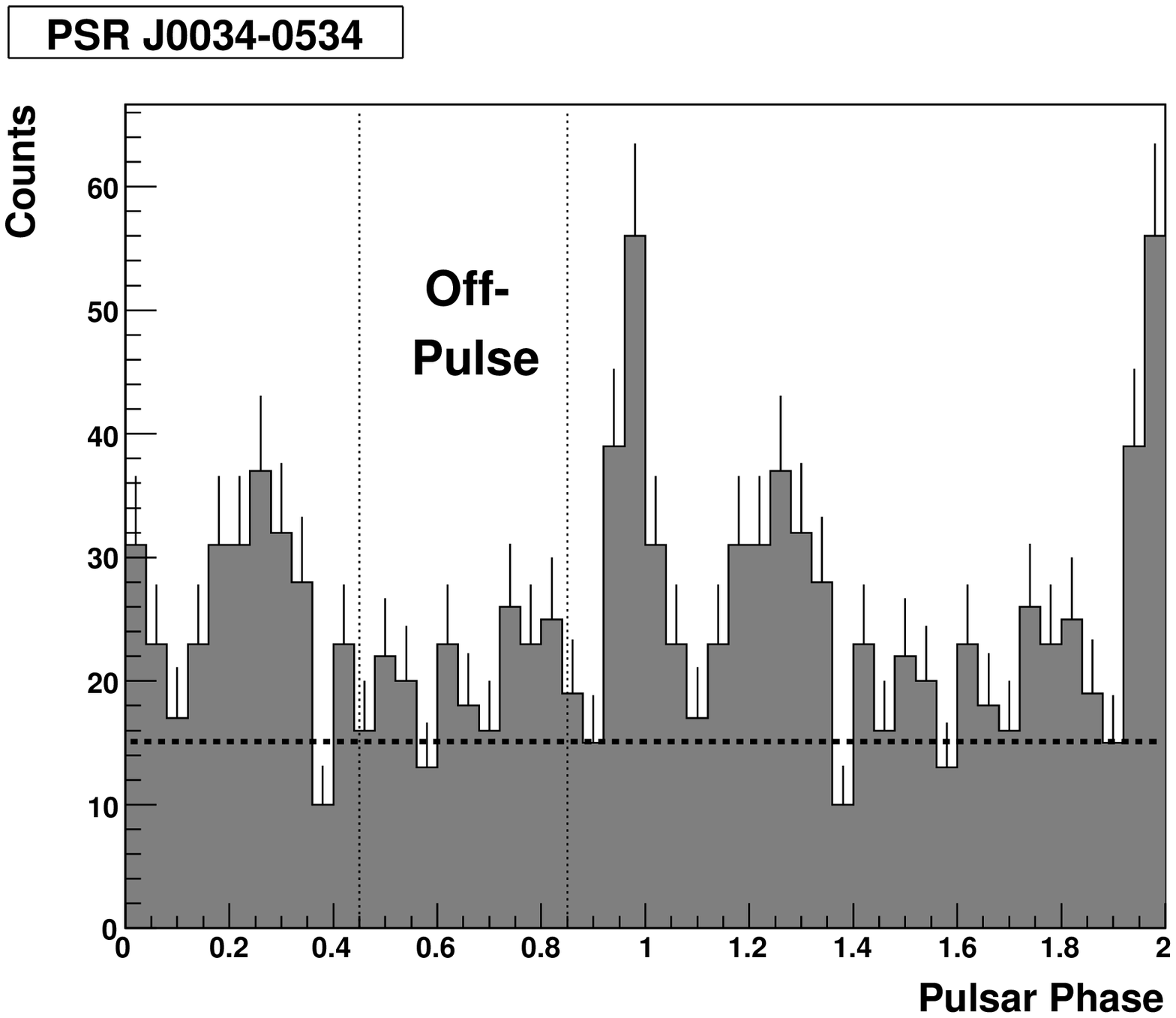}
\end{center}
\end{minipage} \hfill
\begin{minipage}[c]{.45\linewidth}
\begin{center}
\epsscale{1.1}
\plotone{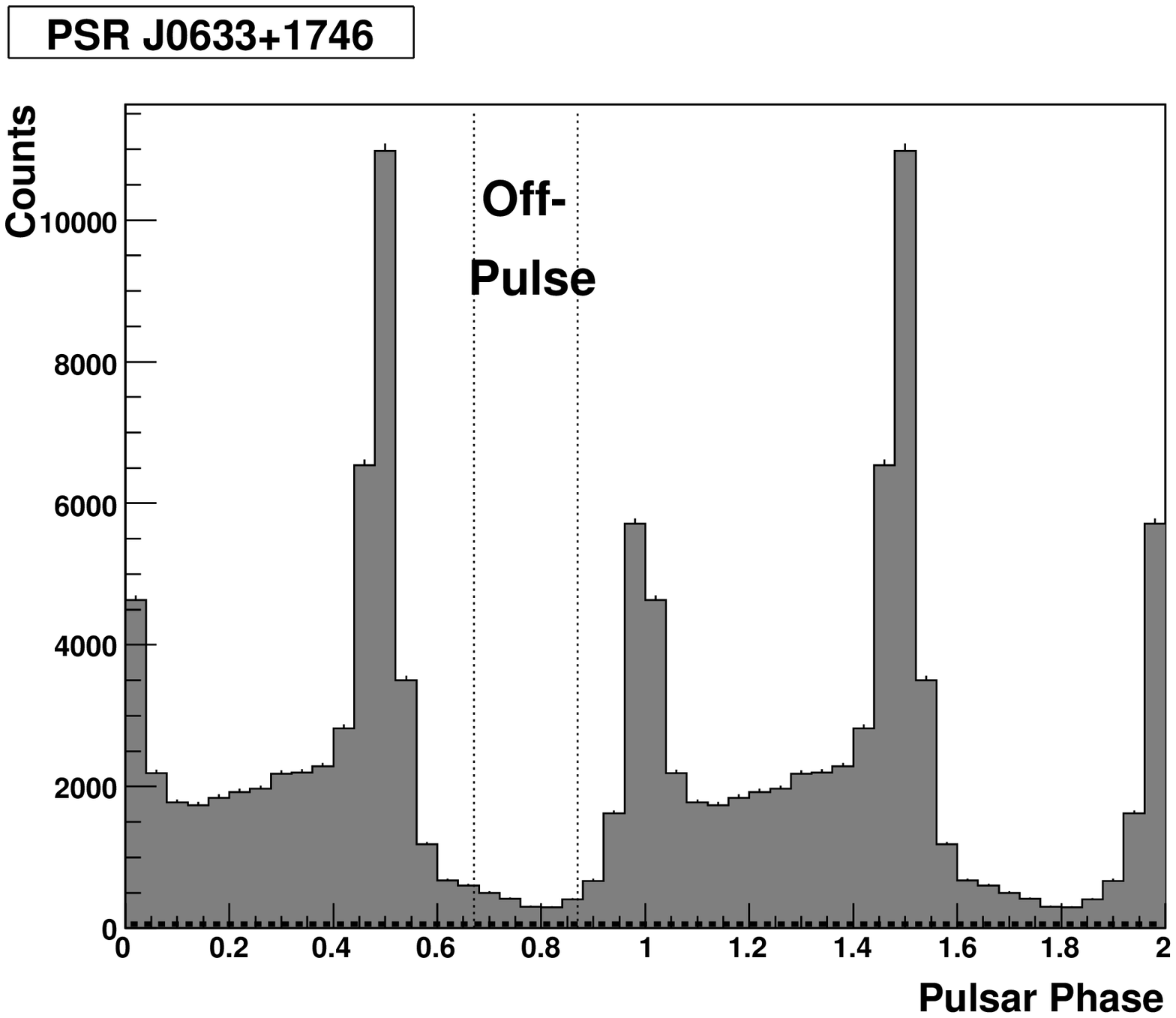}
\end{center}
\end{minipage} \vfill
\begin{minipage}[c]{.45\linewidth}
\begin{center}
\epsscale{1.1}
\plotone{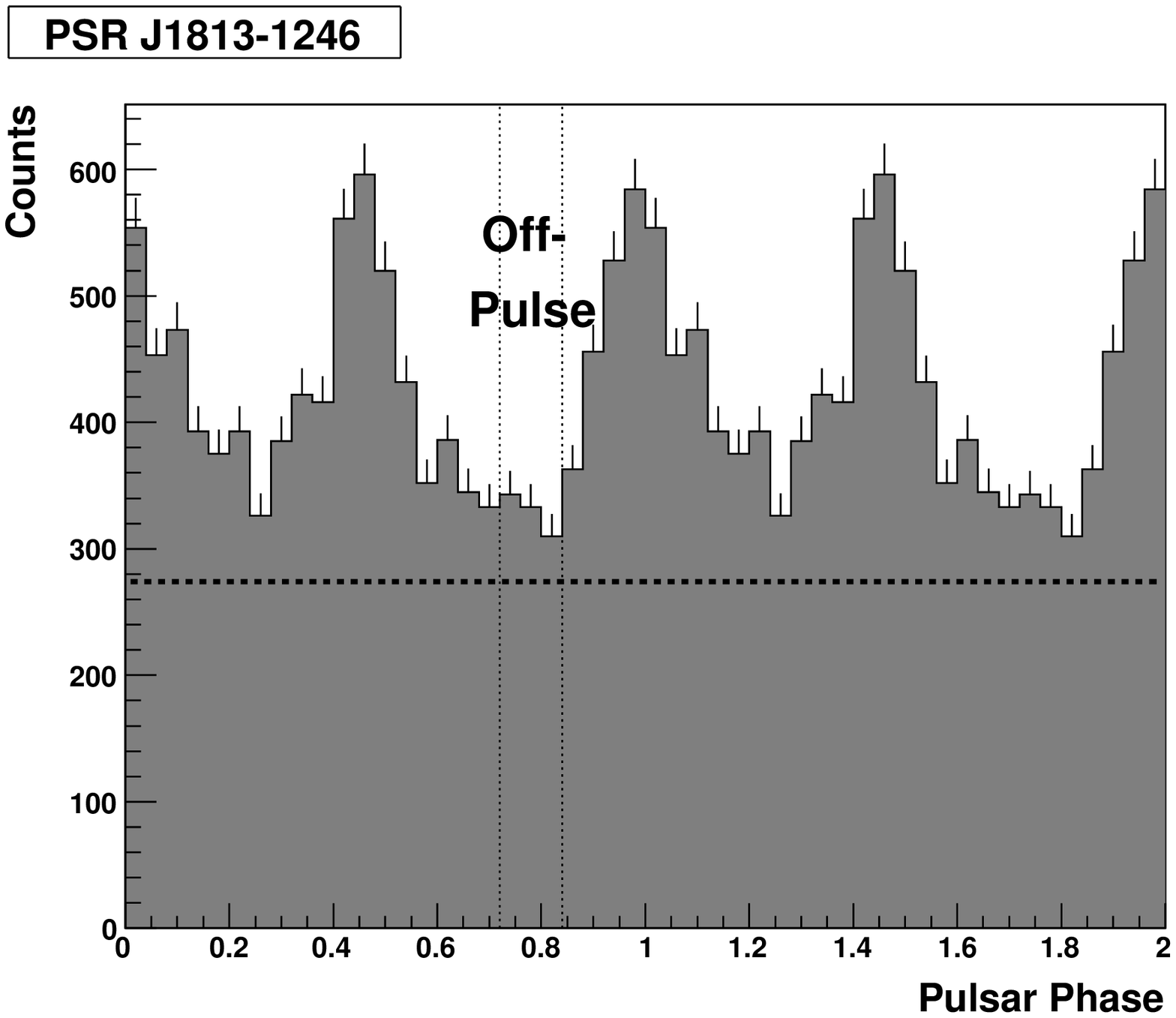}
\end{center}
\end{minipage} \hfill
\begin{minipage}[c]{.45\linewidth}
\begin{center}
\epsscale{1.1}
\plotone{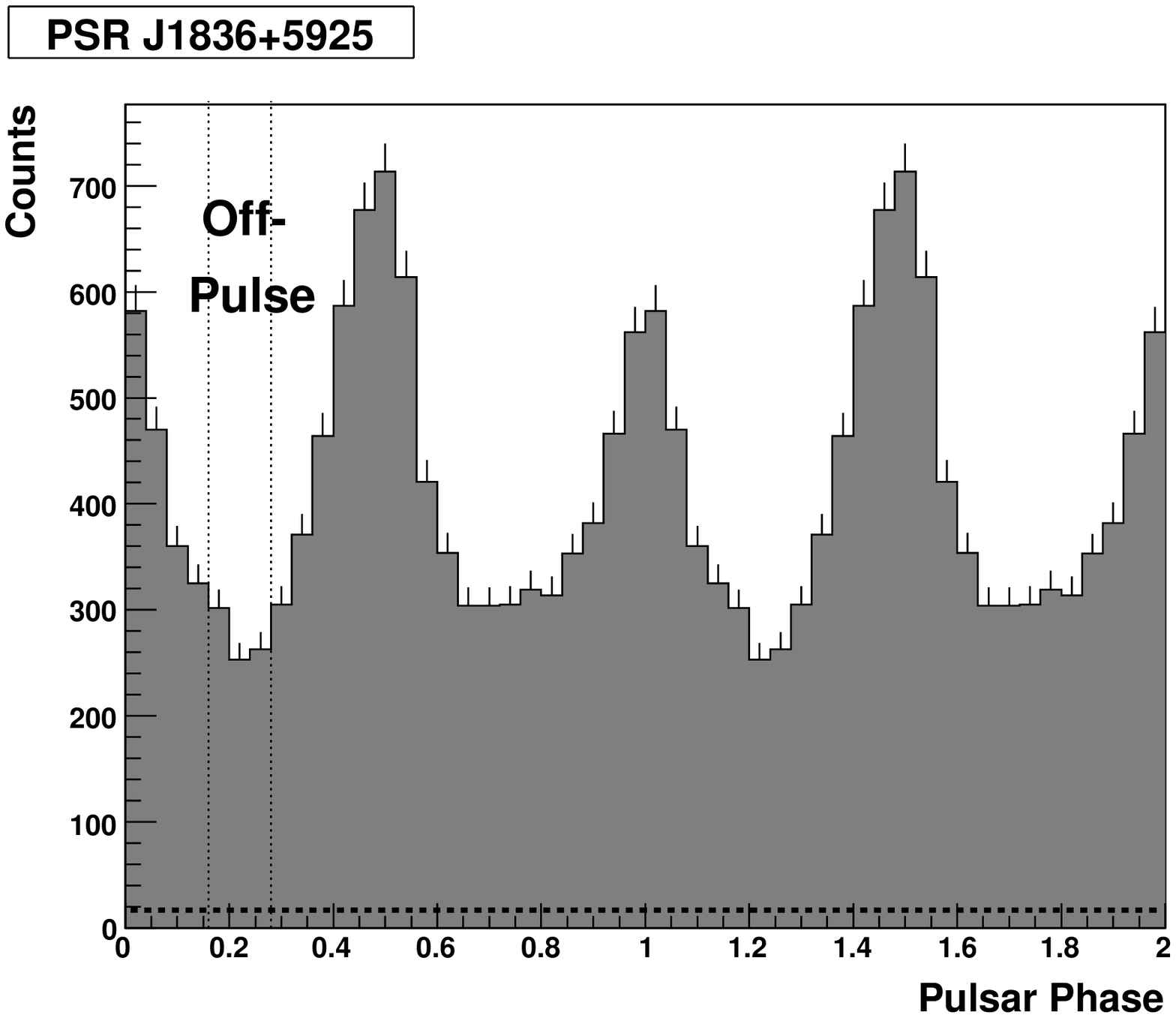}
\end{center}
\end{minipage}
\end{center}
\caption{\label{fig:phaso1}Light curves obtained with photons above 100 MeV in a region of $1^{\circ}$ around J0034$-$0534 (top left), 
J0633+1746 (top right), J1813$-$1246 (bottom left) and J1836+5925 (bottom right). The dashed horizontal line represents the estimated background 
level, as derived from the model used in the spectral fitting. The two dashed vertical lines represent the definition of the off-pulse window, as defined in Table~\ref{tab:def}. Two rotations are shown and 25 bins per rotation.}
\end{figure*}

\begin{figure*}[ht!!]
\begin{center}
\begin{minipage}[c]{.45\linewidth}
\begin{center}
\epsscale{1.1}
\plotone{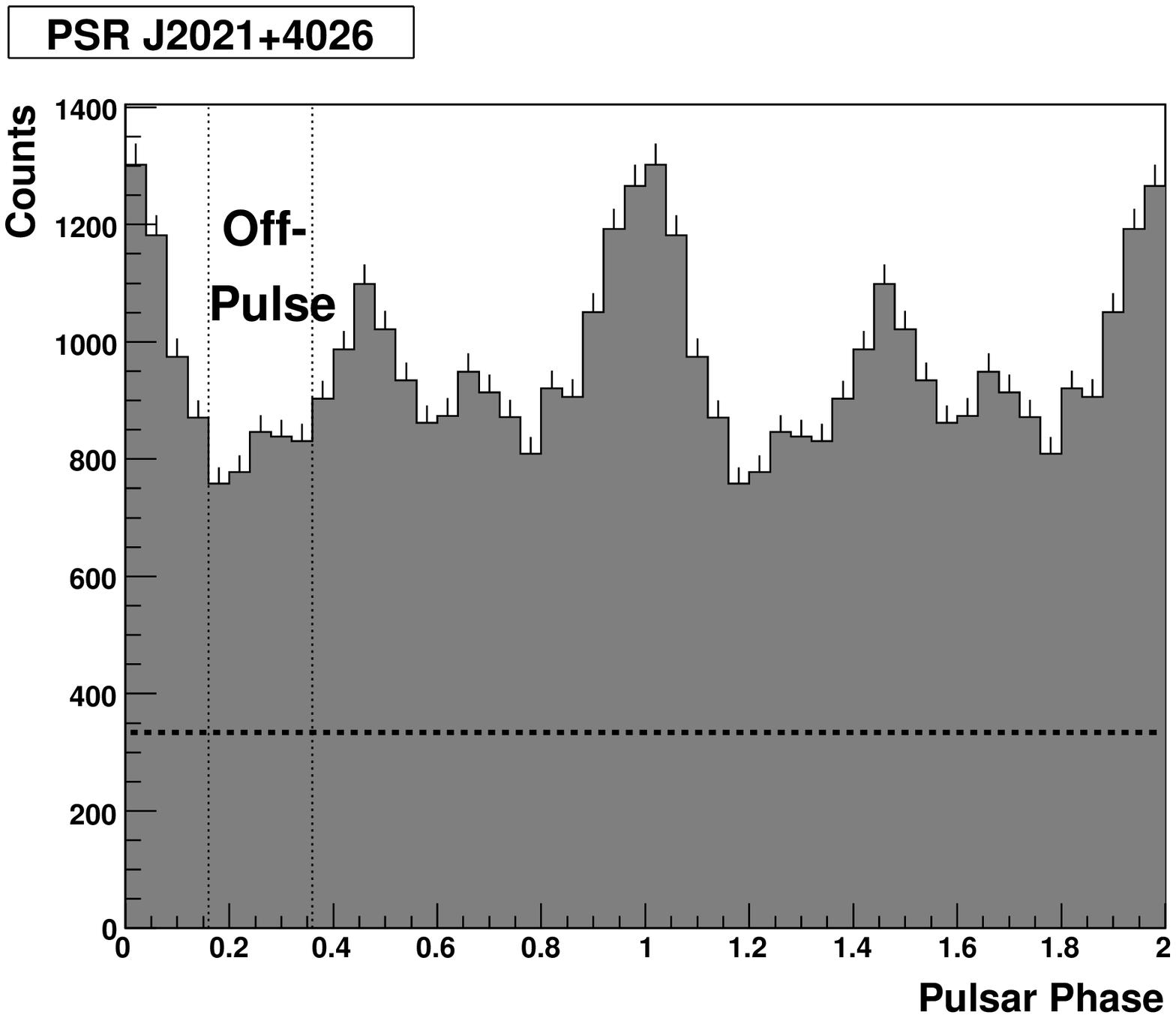}
\end{center}
\end{minipage} \hfill
\begin{minipage}[c]{.45\linewidth}
\begin{center}
\epsscale{1.1}
\plotone{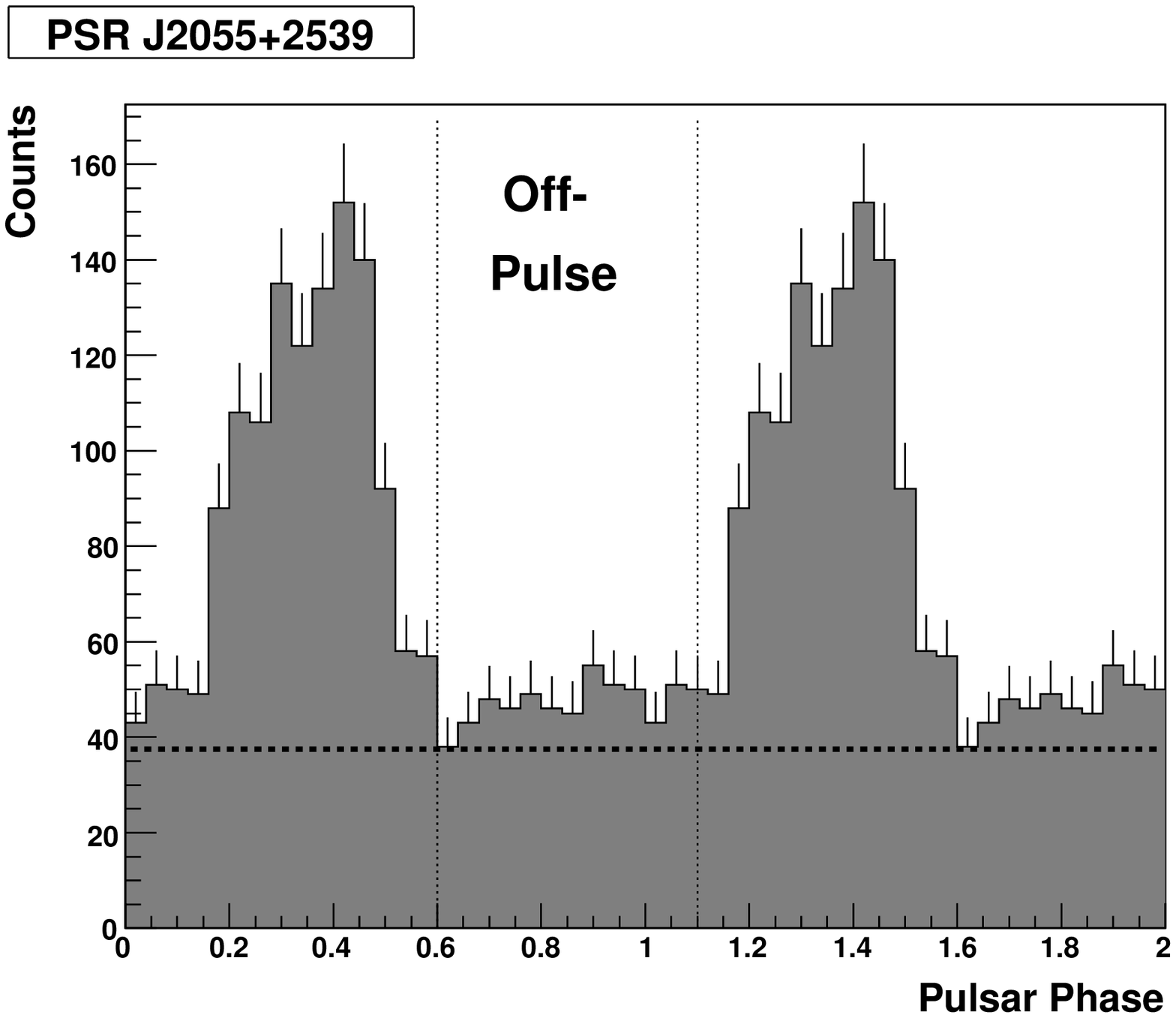}
\end{center}
\end{minipage} \hfill
\begin{minipage}[c]{.45\linewidth}
\begin{center}
\epsscale{1.1}
\plotone{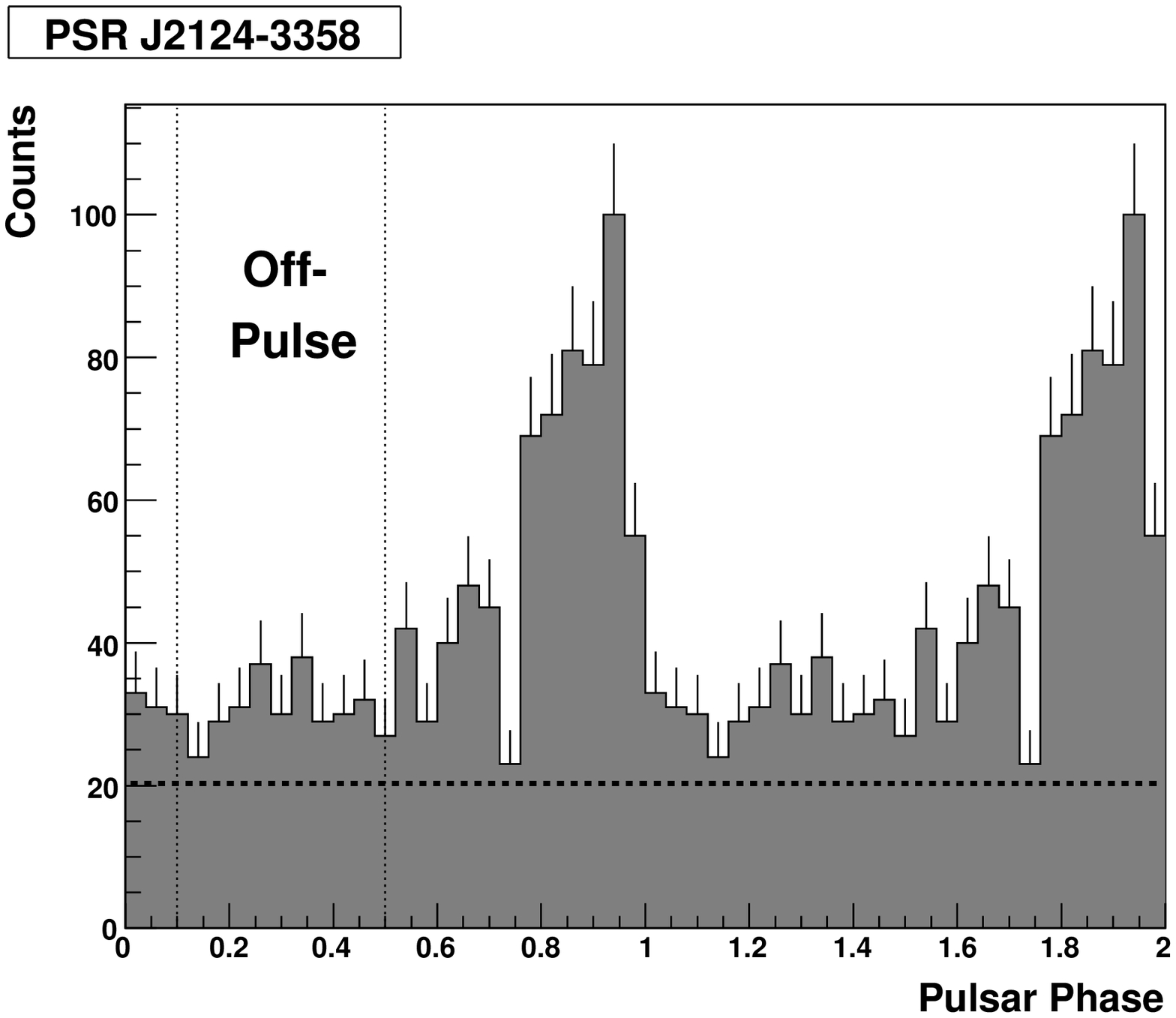}
\end{center}
\end{minipage}
\end{center}
\caption{\label{fig:phaso2}Light curves obtained with photons above 100 MeV in a region of $1^{\circ}$ around J2021+4026 (top left), 
J2055+2539 (top right) and J2124$-$3358 (bottom). Same conventions as for Figure~\ref{fig:phaso1}.
}
\end{figure*}

As can be seen in the light curves presented in Figures~\ref{fig:phaso1} and \ref{fig:phaso2}, the level of off-pulse 
emission of these 7 pulsars greatly varies. The highest levels of off-pulse emission are found for J2021+4026 with $\sim 40$\%, 
J1836+5925 with $\sim 35$\%, J0034$-$0534 and J2124$-$3358, with $\sim 20$\% of the peak heights, while lower levels are found 
for J2055+2539 and J1813$-$1246 with $\sim 10$\% and Geminga with $\sim 5$\% of the peak heights. In the case of TPC models, the highest off-pulse levels in light curves
with two widely-spaced peaks are produced for inclination angle $\alpha > 80^\circ$ and viewing angle $\zeta < 40^\circ$ or 
$\alpha < 40^\circ$ and  $\zeta > 80^\circ$ \citep{Venter2009}. In the case of OG models and wide peak separations, high levels of off-pulse emission 
are produced only for $\alpha > 85^\circ$ and $\zeta < 30^\circ$ \citep{Romani2010}. For both types of model, $\alpha$ and $\zeta$ must be 
very different (i.e., we are viewing the gamma-ray emission at a large angle to the magnetic axis) and
these are precisely the conditions for which our line-of-sight does not cross the radio beam, and for which the pulsar should be radio quiet or 
radio-weak. In fact, all of the non-millisecond pulsars with significant levels of off-pulse emission are radio quiet~\citep[]{8blind,ray2010}. In the case of the 
millisecond pulsars having large polar caps and small magnetospheres, the radio beams are thought to be much larger and a significant fraction 
of the gamma-ray beam size. Therefore, we may still view the radio beams at large angle from the magnetic pole.
The light curve of J0034$-$0534 shows two narrowly-spaced peaks which can be fit in both TPC and OG models for $\alpha = 30^\circ$ and  $\zeta = 70^\circ$, 
but off-pulse emission is predicted in this case only for TPC models \citep{j0034,Venter2010}.
The light curve of J2124$-$3358 has actually been best fit with a PSPC model $\alpha = 40^\circ$ and  $\zeta = 80^\circ$ \citep{Venter2009}, which also predicts 
off-pulse emission. In general, the detection of off-pulse emission in these pulsars constrains the outer gap solutions to a much greater degree 
than for TPC/SG or SL solutions.

\subsection{Constraints on pulsar wind nebulae candidates}
We searched for significant emission in the off-pulse window of 54 gamma-ray pulsars detected by \emph{Fermi}-LAT and 
found only one convincing pulsar wind nebula candidate, J1023-5746 (besides the Crab Nebula and Vela-X). 
However, flux upper limits derived on the steady emission from the nebulae offer new constraints on sources already detected 
in the TeV range (e.g. the PWNe in the Kookaburra complex). Additionally, some PWNe were proposed by~\cite{bednarek} as promising 
sources of $\gamma$-ray emission in the GeV energy range, especially PSR~J0205+6449 and PSR~J2229+6114. We review some 
interesting cases in the following. 

\subsubsection{PSR J0205+6449 and the PWN 3C~58}
\label{mod0205}
The radio source 3C~58 was recognized early to be a supernova remnant (SNR G130.7+3.1) and classified 
as a PWN by~\cite{weiler}. X-ray observations revealed a non-thermal spectrum with the  
photon index becoming steeper toward the outer region of the nebula~\citep{slane04}. 
Flat spectrum radio emission $S_\nu \propto \nu^{-0.12}$ 
covering roughly  $10 \arcmin \times 6 \arcmin$ extends up to $\sim 100$ GHz~\citep{green86,morsi+reich87,salteretal89} and 
corresponds well with infrared \citep{slane08}, and X-ray \citep{slane04} morphologies. Subsequent \emph{Chandra 
X-ray Observatory} observations detected the central pulsar of 3C~58, PSR~J0205+6449. The 
pulsar has a very high spin-down power of $2.7 \times 10^{37}$ erg~s$^{-1}$ and a characteristic age of 
5400~yr. 3C~58 has often been associated with SN~1181~\citep{stephenson}. However, recent investigations of the 
dynamics of the system~\citep{chevalier}, and the velocities of both the radio expansion and optical knots imply 
an age of $\sim 2500$ yr, closer to the characteristic age of PSR~J0205+6449.\\
At TeV energies, both the VERITAS and MAGIC telescopes observed this source and did not find any 
evidence for $\gamma$-ray emission at the position of the pulsar~\citep{magic_2229,pwn_veritas}. 
The upper limits derived from their observations are consistent with the \emph{Fermi} upper limits obtained in the 100 MeV - 100 GeV energy
range of $< 12.9 \times 10^{-12} \, \rm{erg \, cm^{-2} \, s^{-1}}$. This upper limit implies a non-constraining 100 MeV -- 100 GeV
efficiency of $<$~$4 \times 10^{-4}$ -- $6 \times 10^{-4}$ at a distance of 2.6 -- 3.2 kpc.

\subsubsection{PSR~J0633+1746 - Geminga}
The Geminga pulsar is the first representative of a
population of radio-quiet gamma-ray pulsars,
and has been intensely studied since its discovery as
a gamma-ray source by SAS-2, more than thirty years ago
(\citet{fichteletal75}; \citet{kniffenetal75}).
The subsequent ROSAT detection of periodic X-rays from this
source \citep{halpern+holt92} prompted a successful search for
periodicity in high-energy gamma-rays with EGRET \citep{bertschetal92}
X-ray observations with XMM-Newton and Chandra observations indicate a
highly structured pulsar wind nebula
extending $\sim 50 \arcmin$ from the pulsar \citep{pavlovetal10}.

The Milagro Collaboration recently reported a $3.5 \sigma$ source
coincident with Geminga of
extent $\sim 2.6^{\circ}$, likely the result of a PWN \citep{milagro_2229}.
While VHE emission from the vicinity
of this unique pulsar is certainly of interest, the \emph{Fermi} upper
limits on off-pulse emission
above 10 GeV are not very constraining to SED modeling of the single
VHE point.  The lack of data,
compounded by the lack of X-ray data covering the scale of the Milagro
source, renders SED modeling uninformative.

\subsubsection{PWNe in the Kookaburra complex: the Rabbit and K3}
The complex of compact and extended radio sources called Kookaburra~\citep{roberts_kooka} covers nearly one square 
degree along the Galactic plane around $l = 313.4^{\circ}$. This region has been extensively studied to understand 
the nature of the unidentified EGRET source 3EG~J1420-6038~\citep{hartman}. In the North wing of the Kookaburra, 
\cite{damico} discovered the radio pulsar PSR~J1420$-$6048, a young energetic pulsar with period 68~ms, characteristic age 
$\tau_c = 13$~kyr and spin-down power $\dot{E} = 1.0 \times 10^{37}$ erg~s$^{-1}$. Subsequent ASCA and \emph{Chandra} 
observations revealed an X-ray nebula surrounding the pulsar called K3. In the South-Western wing, a bright X-ray emission called the Rabbit has been proposed as a plausible PWN contributing to the $\gamma$-ray signal detected by EGRET. 
At TeV energies, the H.E.S.S. collaboration reported the detection of two bright sources coincident with the Kookaburra complex~\citep{hess_kooka}. 
HESS~J1420$-$607 is centered just North of PSR~J1420$-$6048, with best fit position overlapping the pulsar position. HESS~J1418$-$609 appears to correspond to the Rabbit nebula. However, the gaps in spectral coverage between H.E.S.S. and EGRET did 
not allow a clear statement if the EGRET source is really associated with the X-ray and TeV emission.
Finally, the $\gamma$-ray detection by \emph{Fermi}-LAT of the radio loud pulsar PSR~J1420$-$6048 and the discovery of a radio-quiet PSR~J1418$-$6058, likely powering the Rabbit PWN, brought a new light on this region.\\
In this paper, we searched for $\gamma$-ray emission in the off-pulse of both pulsars, PSR~J1420$-$6048 and PSR~J1418$-$6058, 
and did not detect any significant signal. The 95\% CL upper limits reported in Tables~\ref{tab:res} and \ref{tab:res2} are 
consistent with the prediction made by~\cite{kooka_ave} that do not expect any detection by \emph{Fermi} before 10 yr of 
observations in the most optimistic scenario for PSR~J1420$-$6048. The VHE spectrum of the Rabbit nebula is very similar to that
of K3, so similar emission mechanisms from the Rabbit would correspondingly predict little chance of \emph{Fermi} detection.

\subsubsection{PSR~J1833$-$1034 and G21.5-0.9}
\label{mod1833}
G21.5-0.9 was classified as one of about ten Crab-like SNR and predicted in 1995 to be a gamma-ray 
source~\citep{g21_dejager}. \emph{Chandra} observations revealed the composite nature of the remnant, consisting of a 
centrally peaked PWN and a 4' shell~\citep{bocchino,safi-harbetal01}. The 61.8~ms pulsar PSR~J1833$-$1034 powering the PWN was 
discovered recently through its faint radio emission~\citep{camilo_1833}. With a spin-down power of $\dot{E} = 3.3 \times 10^{37}$ 
erg~s$^{-1}$, PSR~J1833$-$1034 is one of the most energetic pulsars in the Galaxy. INTEGRAL observations revealed that the PWN is also bright 
in the hard X-ray regime~\citep{derosaetal09}. 
At TeV energies, G21.5-0.9 was recently detected by H.E.S.S. and has a point-like nature and a hard spectrum 
($\Gamma$ = 2.08 $\pm$ 0.22)~\citep{hess_1833}. The flux of this source is only 2\% of that of the Crab Nebula. 
Although the spin-down age of this extremely energetic pulsar is 4.6 kyr, VLA measurements of the PWN expansion speed 
place the age at $870^{200}_{-150}$~yr \citep{bandb08}. Kinematic H~{\sc I} and CO distance measurements place G21.5-0.9
some 4.7 kpc distant \citep{tian+leahy08}.  
The lack of detection is consistent with previous modeling undertaken by \citet{dejageretal08}, which predicted GeV gamma-ray flux well below
the \emph{Fermi} upper limits.  

\subsubsection{PSR~J1907+0602 and its TeV PWN}
The TeV source MGRO J1908+06 was discovered by MILAGRO at a median energy of 20 TeV~\citep{abdo_milagro} with a flux $\sim$80\% of the Crab at these energies. 
It was subsequently detected in the 300 GeV--20 TeV range by the H.E.S.S.~\citep{hess_1907} and VERITAS~\citep{veritas_1907} experiments. The \emph{Fermi} 
discovery of the radio-quiet pulsar PSR~J1907+0602~\citep{16blind} within the extent of the TeV source strongly suggests that HESS J1908+063 
is the PWN of PSR~J1907+0602.\\
The upper limits derived on its off-pulse emission (see Table~\ref{tab:res}) are consistent with those previously reported 
in~\cite{fermi_1907} assuming a source extension of $0.3^{\circ}$ and strongly suggest that the spectrum of HESS J1908+063 has a low-energy turnover between 20 GeV and 300 GeV. 
The pulsar is offset from the H.E.S.S. centroid by $15 \arcmin$, and assuming this marks the pulsar birthsite
implies a velocity of $\approx 400 \, {\rm km \, s^{-1}}$ for a distance of 3.2 kpc and an age of 20 kyr.
~\cite{fermi_1907} estimated an upper limit on the 2-10 keV X-ray flux from the pulsar and any arcminute-scale
nebula of $10^{-13} \, {\rm erg \, cm^{-2} \, s^{-1}}$, though given the H.E.S.S. extension is $20 \arcmin$ we 
cannot constrain the magnetic field within the larger nebula.  Therefore without any data outside 
the VHE regime we can only conclude that a cooling break appears around 1 TeV.  Determining whether that break occurs from 
old electrons in a low magnetic field or younger electrons in a high magnetic field requires either an X-ray detection
or upper limit on the extended region.

\subsubsection{PSR J2032+4127 and TeV 2032+4130}
TeV J2032+4130 was the first unclassified TeV source, initially detected by HEGRA \citep{hegra02}, and later confirmed by MAGIC \citep{magic08}.
Subsequent XMM observations revealed a faint diffuse X-ray structure centered on the position of TeV J2032+4130 with the same
$6 \arcmin$ extension as the VHE source \citep{hornsxmm07}. Within the TeV error box \citet{16blind} discovered PSR J2032+4127, 
later confirmed by the radio detection of the pulsar \citep{camiloetal09}, hinting at a PWN origin for the TeV emission.  
The limited radio data and large errors on the XMM X-ray and VHE data prevent precise modeling of all PWN parameters, however, and
the \emph{Fermi} upper limits are approximately an order of magnitude too high to be constraining.  

\subsubsection{PSR J2229+6114 and the Boomerang PWN}
\label{mod2229}
The pulsar PSR J2229+6114 is as young as the Vela pulsar (characteristic age $\tau_c = 10$~kyr), as energetic 
($\dot{E} = 2.2 \times 10^{37}$ erg~s$^{-1}$) and powers the small boomerang-shaped radio and X-ray emitting 
PWN which is part of the supernova remnant G106.3+2.7~\citep{halpern_2229b}. The pulsar distance estimated 
from X-ray absorption is $\sim$3~kpc~\citep{halpern_2229b}, while the dispersion measure used in conjunction 
with the NE2001 model~\citep{cor02} yields a distance of 7.5~kpc~\citep{j2229}; polarization studies and velocity maps of 
HI and CO emission imply a distance of 800~pc~\citep{kothes}. In this paper, we have used the range 0.8--6.5~kpc as reported 
in Table~\ref{tab:def}.

At TeV energies, the MAGIC collaboration placed a constraining point-source upper limit 
of $3.95 \times 10^{-12}$~cm$^{-2}$~s$^{-1}$ at the position of the pulsar~\citep{magic_2229}. 
However, recently, both the MILAGRO and the VERITAS collaboration reported a significant detection in this region. 
The centroid of the extended emission (full angular extent of $0.6^{\circ}$ by $0.4^{\circ}$) detected by VERITAS 
is located $0.4^{\circ}$ away from the pulsar PSR~J2229+6114~\citep{veritas_2229}. The signal detected by MILAGRO 
is spread over a broad $\sim 1^{\circ}$ area encompassing the pulsar position and the main bulk of the remnant, which 
does not allow a definitive association with a particular region of the SNR/pulsar complex~\citep{milagro_2229}.

Using 16~months of \emph{Fermi}-LAT data, we do not report any significant detection at the position of the pulsar 
using its off-pulse photons. The upper limits derived are compatible with the non-detection reported by the MAGIC 
collaboration at TeV energies~\citep{magic_2229}. An offset between the pulsar J2229+6114 
and its PWN is thus required if the emission 
detected at TeV energies by VERITAS and MILAGRO is produced by the pulsar wind nebula. This is the case for several 
PWNe already detected by Cherenkov telescopes, such as HESS~J1825$-$137 and Vela-X, and can be explained by the supernova 
explosion that occured in an inhomogeneous medium leading to an asymmetric reverse shock that displaced the 
PWN~\citep{blondin} towards lower densities. However, the TeV emission detected by VERITAS is coincident with the 
location of molecular clouds which disfavors such a scenario and supports a hadronic origin where $\gamma$-rays 
are produced via proton-proton interactions within the molecular clouds.

\subsection{Population study of pulsar wind nebulae as observed by \emph{Fermi}-LAT}
\label{pop}
\begin{figure*}[ht!!]
\begin{center}
\end{center}
\epsscale{0.9}
\plotone{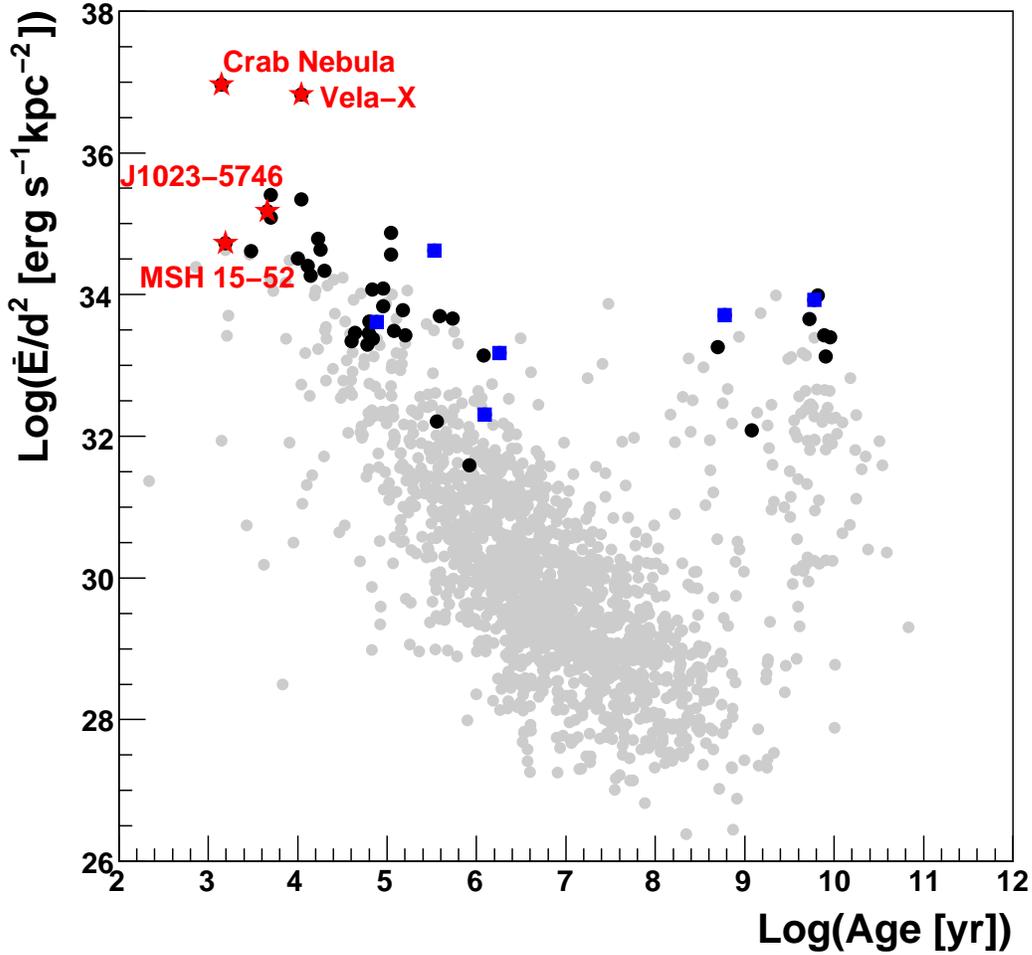}
\caption{\label{fig:pwnage}Spin-down flux at Earth as a function of Age for pulsars in the ATNF catalog; \emph{Fermi}-LAT detected pulsars 
are marked with black circles. Pulsar wind nebulae candidates are marked with red stars; LAT pulsars showing a significant off-pulse emission with a plausible magnetospheric origin are marked with blue squares. For pulsars with a distance range in Table~\ref{tab:def}, we use the geometric mean of the minimum and maximum values. Note that inaccurate distance estimates can introduce artificially low spin-down fluxes, which might account for the handful of pulsar detections below $10^{33}$~erg~s$^{-1}$~kpc$^{-2}$.}
\end{figure*}

\begin{figure*}[ht!!]
\begin{center}
\epsscale{0.9}
\plotone{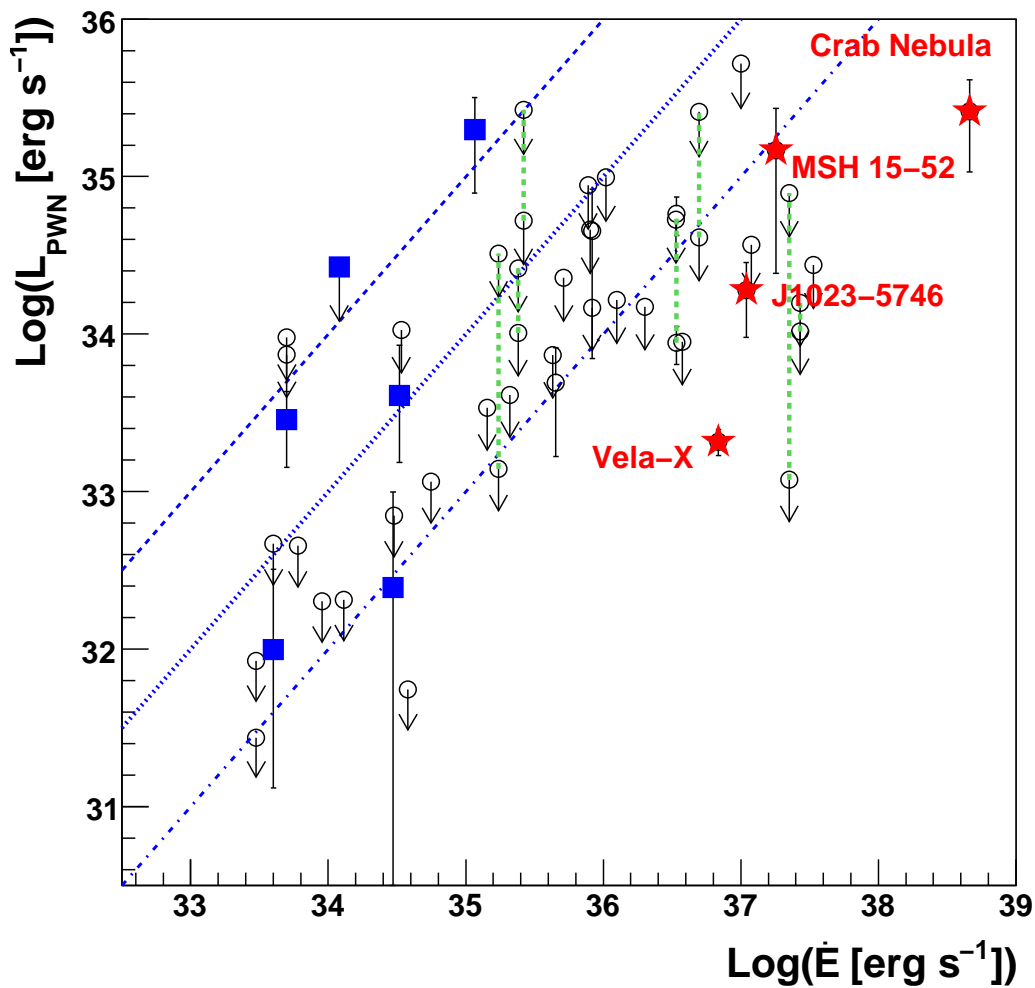}
\end{center}
\caption{\label{fig:pwnpop}
Dependence of the PWN luminosity and the pulsar spin down power for LAT detected pulsars. 
Pulsar wind nebulae candidates are marked with red stars; pulsars showing a significant off-pulse emission with a plausible magnetospheric origin are marked with blue squares. Only pulsars with an estimated distances reported in Table~\ref{tab:def} are plotted. Error bars take into account both the statistical uncertainties on the luminosity and the uncertainty on the distance of the pulsar. Lines correspond to constant $\gamma$-ray 
efficiency: 100\% (dashed), 10\% (dotted), 1\% (dotted-dashed). Pulsars with 2 distance estimates have two markers connected with green dashed 
error bars. The luminosity of the PWN in MSH~15$-$5\emph{2} is taken from~\cite{fermi_msh1552}.}
\end{figure*}

Among the large sample of pulsars analyzed in this paper, two sources are firmly identified as PWNe (the Crab Nebula and Vela-X), one 
source is suggested as a highly plausible candidate (the emission in the off-pulse of PSR~J1023$-$5746), 
in addition to the pulsar wind nebula in MSH~15$-$5\emph{2}~\cite{fermi_msh1552}. As can be seen in Figure~\ref{fig:pwnage}, 
the pulsars powering these PWNe are all young (in the range 1 to 10 kyr) and bright ($\dot{E} \ge 7 \times 10^{36} \, \rm erg s^{-1}$); their associated PWNe are detected by Cherenkov telescopes in the TeV range. Remarkably, these 4 PWNe candidates have a low $\gamma$-ray efficiency in the \emph{Fermi}-LAT energy range, 
$\frac{L_{PWN}}{\dot{E}} < 10^{-2}$, Vela-X, Crab and MSH~15$-$5\emph{2} having even lower $\frac{L_{PWN}}{\dot{E}}$ (see Figure~\ref{fig:pwnpop}). 
This implies that most of the 44 upper limits derived using 16 months of LAT observations are not yet constraining. However for three objects 
(J0659+1414, J1833-1034 and J0205+6449) the upper limits in Figure~\ref{fig:pwnpop} are well below $2 \times 10^{-3} \dot{E}$ suggesting that 
some pulsars are less efficient at producing GeV PWN flux than J1023-5746 and B1509-58. Table~\ref{tab:res} indicates three potential pulsar wind nebulae candidates at $4\sigma$ level: J0007+7303, J1028$-$5819, J1709$-$4429. More data are needed to confirm the detection these 3 sources and significantly probe the variation of the PWN flux with the spin-down power or with the ambient medium.
One can also note two pulsars with significant emission in their off-pulse window and very high efficency: J1836+5925 and J2021+4026. 
These two pulsars show a similar gamma-ray luminosity in the off-pulse and in the on-pulse intervals (renormalized to the total phase interval); 
their luminosities are greater than the spin-down power, assuming a uniform phase-averaged beaming across the sky. However, 
the association distances for the gamma-ray selected pulsars must be treated with caution. 

Reporting the \emph{Fermi} detection of HESS~J1640$-$465, \citet{slane_1640} invoked an excess of 
low energy electrons to account for the GeV data and to explain the continuity in photon spectral index
between the \emph{Fermi} and HESS energy bands.  
While many of the upper limits on PWNe flux catalogued here
are insufficiently constraining to rule out such a situation, we also see no evidence in support of such
an electron spectrum for any object in the catalog.  
We also see no indication of dual electron populations, as proposed to explain Vela-X \citep{velax}. 
The GeV and TeV peaks in Vela-X are comparable in flux, and this would put the GeV flux very near the
\emph{Fermi} sensitivity for a number of Vela-like pulsars with TeV counterparts: PSRs J1418$-$6058, J1420$-$6048, 
J1709$-$4429, and J1907+0602.
Indeed, if most PWNe boast simple electron populations with a power-law index of $\sim 2.3$, 
the majority of these PWNe may prove elusive to \emph{Fermi}.

\acknowledgments
The \textit{Fermi} LAT Collaboration acknowledges generous ongoing support
from a number of agencies and institutes that have supported both the
development and the operation of the LAT as well as scientific data analysis.
These include the National Aeronautics and Space Administration and the
Department of Energy in the United States, the Commissariat \`a l'Energie Atomique
and the Centre National de la Recherche Scientifique / Institut National de Physique
Nucl\'eaire et de Physique des Particules in France, the Agenzia Spaziale Italiana
and the Istituto Nazionale di Fisica Nucleare in Italy, the Ministry of Education,
Culture, Sports, Science and Technology (MEXT), High Energy Accelerator Research
Organization (KEK) and Japan Aerospace Exploration Agency (JAXA) in Japan, and
the K.~A.~Wallenberg Foundation, the Swedish Research Council and the
Swedish National Space Board in Sweden.

Additional support for science analysis during the operations phase is gratefully
acknowledged from the Istituto Nazionale di Astrofisica in Italy and the Centre National d'\'Etudes Spatiales in France.

The Parkes radio telescope is part of the Australia Telescope which is funded by the Commonwealth Government for operation 
as a National Facility managed by CSIRO. We thank our colleagues for their assistance with the radio timing observations. 
The Green Bank Telescope is operated by the National Radio Astronomy Observatory, a facility of the National Science Foundation 
operated under cooperative agreement by Associated Universities, Inc. The Nan\c cay Radio Observatory is operated by the Paris Observatory, 
associated with the French Centre National de la Recherche Scientifique (CNRS). The Lovell Telescope is owned and operated by the University 
of Manchester as part of the Jodrell Bank Centre for Astrophysics with support from the Science and Technology Facilities Council of the 
United Kingdom.

\begin{deluxetable}{lcccc}
\tabletypesize{\scriptsize}
\tablecaption{Observatories, off-pulse definitions and distances of the 54 pulsars analysed
\label{tab:def}}
\tablewidth{0pt}

\tablehead{
\colhead{PSR} & \colhead{ObsID} & \colhead{Off-pulse definition} & \colhead{Distance (kpc)} & \colhead{Observation period rejected (MJD)}
}
\startdata

J0007+7303~$^g$   &   L & 0.4 - 0.8 & 1.4$\pm$0.3 &\\
J0030+0451   &   N & 0.7 - 1.1 & 0.300$\pm$0.090 &\\
J0034-0534   &   N & 0.45 - 0.85 &\\
J0205+6449~$^g$   &   G, J & 0.7 - 1.0 & 2.6--3.2 & 54870 -- 54940 \\
J0218+4232   &   N & 0.9 - 1.1 & 2.5--4 & \\
J0248+6021~$^g$   &  N & 0.7 - 1.1 & 2.0$\pm0.2$ & 55161 -- 55181 \\
J0357+32     &   L & 0.35 - 0.85 & \\
J0437$-$4715   & P & 0.7 - 1.2 & 0.1563$\pm$0.0013 & \\
J0534+2200   &   N, J & 0.5 - 0.85 & 2.0$\pm$0.5 & \\
J0613$-$0200   & N & 0.6 - 1.05 & 0.48$^{+0.19}_{-0.11}$ & \\
J0631+1036~$^g$   &  N, J & 0.9 - 1.15 & 0.75--3.62 & \\
J0633+0632   &  L & 0.6 - 0.8 & \\
J0633+1746   &  L & 0.67 - 0.87 & 0.250$^{+0.120}_{-0.062}$ & \\
J0659+1414   &  N, J & 0.45 - 1.0 & 0.288$^{ +0.033}_{-0.027}$ & \\
J0742$-$2822~$^g$   & N, J & 0.8 - 1.4 & 2.07$^{+1.38}_{-1.07}$ & \\
J0751+1807   & N & 0.7 - 1.05 & 0.6$^{ +0.6}_{-0.2}$ & \\
J0835$-$4510   & P & 0.7 - 1.0 & 0.287$^{+0.019}_{-0.017}$ & \\
J1023$-$5746~$^g$   & L & 0.85 - 1.13 & 2.4 & \\
J1028$-$5819   & P & 0.8 - 1.05  & 2.33$\pm$0.70 & \\
J1044$-$5737   & L & 0.75 - 1.1 & 1.5 & \\
J1048$-$5832   & P & 0.7 - 1.05 & 2.71$\pm$0.81 & \\
J1057$-$5226   & P & 0.7 - 0.2 & 0.72 $\pm$0.2 & \\
J1124$-$5916~$^g$   & L & 0.92 - 0.08 & 4.8$^{+0.7}_{-1.2 }$ & \\
J1413$-$6205~$^g$   & L & 0.7 - 0.15 & 1.4 & 54682 -- 54743 \\
J1418$-$6058   & L & 0.55 - 0.90 & 2--5 & \\
J1420$-$6048   & P & 0.6 - 1.1 & 5.6$\pm$1.7 & \\
J1429$-$5911   & L & 0.85 - 0.1 & 1.6 & \\
J1459$-$60     & L & 0.34 - 0.69 & & \\
J1509$-$5850   & P & 0.6 - 1.0 & 2.6$\pm$0.8 & \\
J1614$-$2230   & G & 0.92 - 1.14 & 1.27$\pm$0.39 & \\
J1709$-$4429~$^g$   & P & 0.65 - 1.1 & 1.4--3.6 & \\
J1718$-$3825   & N, P & 0.65 - 1.15 & 3.82$\pm$1.15 & \\
J1732$-$31     & L & 0.54 - 0.89 &  & \\
J1741$-$2054   & L & 0.67 - 1.18 & 0.38$\pm$0.11 & \\
J1744$-$1134   & N & 0.15 - 0.35 & 0.357$^{+0.043}_{-0.035}$ & \\
J1809$-$2332   & L & 0.45 - 0.85 & 1.7$\pm$1.0 & \\
J1813$-$1246~$^g$   & L & 0.72 - 0.84 &  & 55084 -- 55181 \\
J1826$-$1256   & L & 0.60 - 0.90 & & \\
J1833$-$1034   & G & 0.75 - 1.1 & 4.7$\pm$0.4 & \\
J1836+5925   & L & 0.16 - 0.28 & $<$0.8 & \\
J1846+0919   & L & 0.65 - 1.0 & 1.2 & \\
J1907+06     & L & 0.51 - 0.91 &  &  \\
J1952+3252   & J, N & 0.7 - 1.0 & 2.0$\pm$0.5 & \\
J1954+2836   & L & 0.85 - 0.2 & 1.7 & \\
J1957+5033   & L & 0.6 - 0.05 & 0.9 & \\
J1958+2846   & L & 0.55 - 0.90 &  & \\
J2021+3651~$^g$   & G & 0.75 - 1.05 & 2.1$^{+2.1}_{-1.0}$ & \\
J2021+4026   & L & 0.16 - 0.36 & 1.5$\pm$0.45 &  \\
J2032+4127   & L & 0.30 - 0.45 \& 0.90 - 0.05 & 1.6--3.6 &  \\
J2043+2740   & N, J & 0.68 - 0.08 & 1.80$\pm$0.54 & \\
J2055+2539   & L & 0.6 - 0.1 & 0.4 & \\
J2124$-$3358   & N & 0.1 - 0.5 & 0.25$^{+0.25}_{-0.08}$ &  \\
J2229+6114~$^g$   & G, J & 0.68 - 1.08 & 0.8--6.5 & \\
J2238+59     & L &  0.65 - 0.90  \\
\enddata

\tablecomments{
Column 1 lists the pulsars; a ``$g$'' indicates that one or several glitches occured during the observation period. For some pulsars, 
these glitches led to restrict the dataset to avoid any contamination of pulsed emission during the glitch: the observation period rejected in these cases is indicated in column 5 (Modified Julian Day).
Column 2 indicates the observatories that provided ephemerides:
``G'' -- Green Bank Telescope;
``J'' -- Lovell telescope at Jodrell Bank;
``L'' -- Large Area Telescope;
``N'' -- Nan\c cay Radio Telescope;
``P'' -- Parkes radio telescope.
Column 3 lists the off-pulse phase range used in the spectral analysis. Column 4 presents the best known distances of 54 the pulsars analyzed in this paper.}

\end{deluxetable}

\begin{landscape}
\begin{deluxetable}{lcccrc}
\tabletypesize{\scriptsize}
\tablecaption{Spectral fit results for 54 LAT-detected pulsars
\label{tab:res}}
\tablewidth{0pt}

\tablehead{
\colhead{PSR} & \colhead{TS} & \colhead{$F_{0.1-100}$} & \colhead{$G_{0.1-100}$} & \colhead{$\Gamma$} & \colhead{Luminosity}\\
\colhead{ } & \colhead{ } & \colhead{($\rm 10^{-9} \,ph \ cm^{-2}\, s^{-1}$)} & \colhead{($\rm 10^{-12} \,erg \ cm^{-2}\, s^{-1}$)} & \colhead{ }  & \colhead{($\rm 10^{33} \,erg s^{-1}$)}
}
\startdata

J0007+7303   & 24.3 & $<$63.23 & $<$69.94 &  & $<$16.40 \\
J0030+0451   & 3.4 & $<$7.07 & $<$7.83 &  & $<$0.08 \\
J0034-0534   & 29.1 & 17.26 $\pm$ 5.70 & 11.09 $\pm$ 2.68 & 2.27 $\pm$ 0.17 & 0.25$_{-0.25}^{+0.75}$ \\
J0205+6449   & 1.3 & $<$11.63 & $<$12.88 &  & $<$10.42 -- 15.78 \\
J0218+4232   & 1.2 & $<$12.33 & $<$13.65 &  & $<$10.21 -- 26.13 \\
J0248+6021   & 0.3 & $<$7.77 & $<$8.59 & & $<$4.11 \\
J0357+32     & 0.0 & $<$3.94 & $<$4.36 &  & \nodata \\
J0437$-$4715 & 10.7 & $<$8.50 & $<$9.41 &  & $<$0.03 \\
J0534+2200$^a$   & 2775.6 & 980.00 $\pm$ 70.00 & 540.92 $\pm$ 46.73 & 2.15 $\pm$ 0.03 & 258.88$\pm 151.81$ \\
J0613$-$0200 & 4.0 & $<$6.74 & $<$7.46 &  & $<$0.21 \\
J0631+1036   & 2.5 & $<$18.72 & $<$20.72 &  & $<$1.39 -- 32.49 \\
J0633+0632   & 6.3 & $<$32.50 & $<$35.97 &  & \nodata \\
J0633+1746   & 5101.2 & 1115.54 $\pm$ 32.31 & 749.44 $\pm$ 22.24 & 2.24 $\pm$ 0.02 & 4.07$_{-2.53}^{+4.42}$ \\
J0659+1414   & 0.6 & $<$5.04 & $<$5.58 &  & $<$0.05 \\
J0742$-$2822 & 0.0 & $<$5.99 & $<$6.63 &  & $<$3.40 \\
J0751+1807   & 6.4 & $<$9.52 & $<$10.53 & & $<$0.45 \\
J0835$-$4510$^b$ & 284.3 & 405.44 $\pm$ 26.75 & 210.25 $\pm$ 13.87 & 2.30 $\pm$ 0.10 & 2.07$^{+0.41}_{-0.38}$ \\
J1023$-$5746 & 25.1 & 1.33 $\pm$ 1.14 & 27.58 $\pm$ 13.73 & 1.05 $\pm$ 0.36 & 19.01 $\pm$ 9.46 \\
J1028$-$5819 & 20.8 & $<$88.79 & $<$98.27 &  & $<$63.83 \\
J1044$-$5737 & 0.0 & $<$11.93 & $<$13.20 &  & $<$46.05\\
J1048$-$5832 & 0.0 & $<$15.33 & $<$16.96 & & $<$14.90  \\
J1057$-$5226 & 1.2 & $<$10.26 & $<$11.35 &  & $<$0.70 \\
J1124$-$5916 & 0.0 & $<$12.09 & $<$13.38 & & $<$36.88 \\
J1413$-$6205 & 5.3 & $<$4.83 & $<$5.34 &  & $<$14.72 \\
J1418$-$6058 & 10.3 & $<$77.73 & $<$86.03 & & $<$41.17 -- 257.33 \\
J1420$-$6048 & 3.0 & $<$125.98 & $<$139.42 & & $<$523.13 \\
J1429$-$5911 & 0.0 & $<$21.26 & $<$23.53 &  & $<$88.29 \\
J1459$-$60   & 1.2 & $<$23.17 & $<$25.64 &  & \nodata \\
J1509$-$5850 & 1.1 & $<$25.47 & $<$28.19 &  & $<$22.80 \\
J1614$-$2230 & 5.3 & $<$22.01 & $<$24.36 &  & $<$9.55 \\
J1709$-$4429 & 16.5 & $<$35.59 & $<$39.39 & & $<$61.08 \\
J1718$-$3825 & 0.0 & $<$8.53 & $<$9.44 &  & $<$16.48  \\
J1732$-$31   & 0.0 & $<$7.40 & $<$8.19 &  & \nodata \\
J1741$-$2054 & 0.3 & $<$10.52 & $<$11.64 &  & $<$0.20  \\
J1744$-$1134 & 6.9 & $<$27.68 & $<$30.64 &  & $<$0.47 \\
J1809$-$2332 & 1.9 & $<$19.19 & $<$21.25 &  & $<$7.35 \\
J1813$-$1246 & 38.1 & 295.55 $\pm$ 23.44 & 119.03 $\pm$ 9.29 & 2.65 $\pm$ 0.14 & \nodata \\
J1826$-$1256 & 9.7 & $<$145.17 & $<$160.67 & & \nodata \\
J1833$-$1034 & 0.0 & $<$9.38 & $<$10.38 & & $<$27.43 \\
J1836+5925   & 2293.6 & 579.60 $\pm$ 28.56 & 542.16 $\pm$ 34.03 & 2.07 $\pm$ 0.03 & 26.77$\pm 1.23$\\
J1846+0919   & 0.0 & $<$4.79 & $<$5.30 &  & $<$10.66 \\
J1907+06     & 0.7 & $<$17.02 & $<$18.83 &  & \nodata \\
J1952+3252   & 2.4 & $<$16.88 & $<$18.68 &  & $<$8.94 \\
J1954+2836   & 2.4 & $<$21.49 & $<$23.78 &  & $<$99.04 \\
J1957+5033   & 0.3 & $<$5.45 & $<$6.04 &  & $<$7.40 \\
J1958+2846   & 2.6 & $<$15.43 & $<$17.07 &  & \nodata \\
J2021+3651   & 15.8 & $<$91.48 & $<$101.24 &  & $<$53.42 \\
J2021+4026   & 2229.1 & 1603.0 $\pm$ 11.2 & 888.12 $\pm$ 8.56 & 2.36 $\pm$ 0.02  & 198.45$\pm 119.83$\\
J2032+4127   & 1.2 & $<$154.91 & $<$171.45 & & $<$52.51 -- 265.86\\
J2043+2740   & 0.0 & $<$2.71 & $<$2.99 &  & $<$1.16 \\
J2055+2539   & 36.7 & 38.41 $\pm$ 10.10 & 17.59 $\pm$ 3.34 & 2.51 $\pm$ 0.15 & 2.87 $\pm 1.44$\\
J2124$-$3358 & 64.6 & 22.78 $\pm$ 6.43 & 21.81 $\pm$ 4.44 & 2.06 $\pm$ 0.14  & 0.10$_{-0.09}^{+0.22}$\\
J2229+6114   & 0.4 & $<$14.05 & $<$15.55 &  & $<$1.19 -- 78.61 \\
J2238+59     & 0.0 & $<$14.92 & $<$165.10 &  & \nodata \\
\enddata

\tablenotetext{\textit{a}}{The spectral parameters of the Crab Nebula are derived using \citet{crab}.}
\tablenotetext{\textit{b}}{The spectral parameters are derived assuming a uniform disk morphology as described in \citet{velax}.}

\tablecomments{Results of the maximum likelihood spectral fits for the off-pulse emission of LAT gamma-ray 
pulsars (see Section~\ref{ana}) between 100 MeV and 100 GeV. Pulsar wind nebula spectra are fitted with a power-law model 
(photon index $\Gamma$, photon flux $F$ and energy flux $G$) assuming a point-source at the position of the pulsar.  
The test statistic ($TS$) for the source significance is provided in Column 2, the photon flux $F$ and the energy flux are reported in Columns 3 and 4 while the photon index is listed in Column 5 when $TS \ge$ 25. The photon flux and energy flux obtained from the likelihood analysis are replaced by a 2$\sigma$ upper limit when $TS <$ 25 (assuming a photon index $\Gamma$=2). The total gamma-ray luminosity $L_\gamma$ is listed in column 6. The error 
on the photon flux and the photon index only include statistical uncertainties while the error on $L_\gamma$ include the statistical 
uncertainties on the flux and the distance uncertainties.}

\end{deluxetable}
\end{landscape}

\begin{landscape}
\begin{deluxetable}{lccrccrccr}
\tabletypesize{\scriptsize}
\tablecaption{Spectral fit results for 54 LAT-detected pulsars
\label{tab:res2}}
\tablewidth{0pt}

\tablehead{
\multicolumn{1}{|c|}{PSR} & \multicolumn{3}{c|}{0.1 - 1 GeV} & \multicolumn{3}{c|}{1 - 10 GeV} & \multicolumn{3}{c|}{10 - 100 GeV}   \\
\cline{2-10}
\multicolumn{1}{|c|}{ } & \colhead{TS} & \colhead{$F_{0.1-1}$} & \colhead{$\Gamma$} & \colhead{TS} & \colhead{$F_{1-10}$} & \colhead{$\Gamma$} & \colhead{TS} & \colhead{$F_{10-100}$} & \colhead{$\Gamma$}\\
\multicolumn{1}{|c|}{ } &  & \colhead{($\rm 10^{-9} \,ph \ cm^{-2}\, s^{-1}$)} &  & & \colhead{($\rm 10^{-9} \,ph \ cm^{-2}\, s^{-1}$)} &  &  & \colhead{($\rm 10^{-9} \,ph \ cm^{-2}\, s^{-1}$)} &  
}
\startdata

J0007+7303   & 22.8 & $<$54.39 &  & 9.2 & $<$2.75 & & 0.0 & $<$0.24 & \\
J0030+0451   & 6.3 & $<$17.11 & & 3.7 & $<$0.67 & & 0.0 & $<$0.18 & \\
J0034-0534   & 26.9 & 12.36 $\pm$ 7.15 & 1.47 $\pm$ 0.60 & 16.5 & 0.97 $\pm$ 0.35 & 3.14 $\pm$ 0.78 & 0.0 & $<$0.17 & \\
J0205+6449   & 0.0 & $<$12.35 & & 2.5 & $<$1.99 & & 1.6 & $<$0.30 & \\
J0218+4232   & 0.3 & $<$24.46 & & 1.1 & $<$1.39 & & 0.0 & $<$0.47 & \\
J0248+6021   & 0.0 & $<$5.40 & & 0.0 & $<$0.35 & & 0.3 & $<$0.25 & \\
J0357+32     & 0.0 & $<$9.15 & & 0.0 & $<$0.56 & & 0.0 & $<$0.14 & \\
J0437$-$4715 & 4.1 & $<$13.83 & & 8.7 & $<$0.94 & & 0.0 & $<$0.18 & \\
J0534+2200$^a$   & 1054.5 & 785.14 $\pm$ 45.37 & 3.20 $\pm$ 0.07 & 1206.9 & 22.93 $\pm$ 1.44 & 1.59 $\pm$ 0.10 & 830.7 & 5.12 $\pm$ 0.56 & 1.91 $\pm$ 0.19 \\
J0613$-$0200 & 5.5 & $<$26.97 & & 0.5 & $<$0.72 & & 0.0 & $<$0.16 & \\
J0631+1036   & 4.6 & $<$49.87 & & 0.27 & $<$0.18 & & 0.0 & $<$1.11 & \\
J0633+0632   & 7.4 & $<$67.47 & & 1.3 & $<$2.85 & & 0.0 & $<$0.42 & \\
J0633+1746   & 3377.3 & 837.66 $\pm$ 32.20 & 1.81 $\pm$ 0.05 & 2028.4 & 65.41 $\pm$ 3.08 & 3.26 $\pm$ 0.11 & 0.0 & $<$0.35 & \\
J0659+1414   & 0.5 & $<$11.94 & & 1.7 & $<$0.59 & & 0.0 & $<$0.13 & \\
J0742$-$2822 & 0.0 & $<$24.54 & & 0.0 & $<$0.94 & & 0.0 & $<$0.13 & \\
J0751+1807   & 1.7 & $<$13.99 & & 11.4 & $<$1.46 & & 0.0 & $<$0.19 & \\
J0835$-$4510$^b$ & 199.8 & 329.76 $\pm$ 34.54 & 2.15 $\pm$ 0.11 & 97.9 & 18.36 $\pm$ 2.33 & 2.22 $\pm$ 0.20 & 2.4 & $<$0.89 & \\
J1023$-$5746 & 0.0 & $<$12.55 & & 0.9 & $<$2.42 & & 17.2 & 0.46 $\pm$ 0.22 & 1.02 $\pm$ 0.73 \\
J1028$-$5819 & 15.8 & $<$180.07 & & 15.0 & $<$7.18 & & 0.0 & $<$0.53 & \\
J1044$-$5737 & 0.6 & $<$45.53 & & 0.0 & $<$1.46 & & 0.0 & $<$0.36 & \\
J1048$-$5832 & 0.0 & $<$40.52 & & 0.4 & $<$2.12 & & 0.0 & $<$0.33 & \\
J1057$-$5226 & 0.8 & $<$22.44 & & 2.4 & $<$1.33 & & 0.0 & $<$0.21 & \\
J1124$-$5916 & 0.0 & $<$49.42 & & 0.2 & $<$2.52 & & 0.0 & $<$0.49 & \\
J1413$-$6205 & 2.8 & $<$55.63 & & 11.8 & $<$6.77 & & 0.0 & $<$0.37 & \\
J1418$-$6058 & 5.2 & $<$94.24 & & 6.9 & $<$10.28 & & 1.1 & $<$0.66 & \\
J1420$-$6048 & 6.7 & $<$291.45 & & 0.0 & $<$9.62 & & 1.1 & $<$0.83 & \\
J1429$-$5911 & 0.0 & $<$54.23 & & 0.0 & $<$3.25 & & 0.0 & $<$0.48 & \\
J1459$-$60   & 5.0 & $<$68.17 & & 0.1 & $<$1.76 & & 1.5 & $<$0.56 & \\
J1509$-$5850 & 0.7 & $<$65.47 & & 0.4 & $<$3.33 & & 0.0 & $<$0.33 & \\
J1614$-$2230 & 2.5 & $<$34.37 & & 7.7 & $<$2.30 & & 0.0 & $<$0.43 & \\
J1709$-$4429 & 15.5 & $<$96.67 & & 3.3 & $<$2.36 & & 0.0 & $<$0.22 & \\
J1718$-$3825 & 0.5 & $<$56.84 & & 0.0 & $<$0.92 & & 0.1 & $<$0.24 & \\
J1732$-$31   & 0.0 & $<$35.79 & & 0.0 & $<$1.09 & & 0.0 & $<$0.25 & \\
J1741$-$2054 & 0.9 & $<$27.77 & & 4.4 & $<$1.40 & & 0.0 & $<$0.14 & \\
J1744$-$1134 & 10.3 & $<$71.71 & & 4.9 & $<$1.92 & & 0.0 & $<$0.45 & \\
J1809$-$2332 & 2.8 & $<$50.50 & & 8.8 & $<$2.40 & & 1.2 & $<$0.27 & \\
J1813$-$1246 & 32.7 & 261.21 $\pm$ 73.15 & 2.25 $\pm$ 0.27 & 16.3 & 8.43 $\pm$ 2.57 & 3.05 $\pm$ 0.69 & 3.6 & $<$1.09 & \\
J1826$-$1256 & 9.5 & $<$251.30 & & 3.4 & $<$5.81 & & 0.1 & $<$0.38 & \\
J1833$-$1034 & 0.0 & $<$13.33 & & 0.1 & $<$1.67 & & 0.2 & $<$0.40 & \\
J1836+5925   & 1381.5 & 401.84 $\pm$ 27.39 & 1.56 $\pm$ 0.09 & 1014.1 & 51.36 $\pm$ 3.89 & 2.93 $\pm$ 0.16 & 0.0 & $<$0.74 & \\
J1846+0919   & 0.0 & $<$17.79 & & 0.0 & $<$0.61 & & 0.0 & $<$0.19 & \\
J1907+06     & 1.9 & $<$70.95 & & 0.9 & $<$2.33 & & 0.0 & $<$0.20 & \\
J1952+3252   & 1.4 & $<$50.81 & & 1.3 & $<$1.58 & & 0.0 & $<$0.23 & \\
J1954+2836   & 1.4 & $<$47.98 & & 2.6 & $<$2.55 & & 0.0 & $<$0.25 & \\
J1957+5033   & 1.2 & $<$16.35 & & 0.2 & $<$0.57 & & 0.0 & $<$0.17 & \\
J1958+2846   & 0.0 & $<$27.14 & & 5.3 & $<$1.78 & & 0.0 & $<$0.26 & \\
J2021+3651   & 17.7 & 85.90 $\pm$ 30.02 & 1.90 $\pm$ 0.31 & 13.2 & $<$4.50 & & 0.0 & $<$0.25 & \\
J2021+4026   & 1718.2 & 1344.75 $\pm$ 55.56 & 2.03 $\pm$ 0.05 & 936.2 & 73.76 $\pm$ 3.93 & 3.04 $\pm$ 0.11 & 12.16 & $<$1.24 & \\
J2032+4127   & 3.5 & $<$133.39 & & 0.0 & $<$2.08 & & 1.3 & $<$0.56 & \\
J2043+2740   & 0.0 & $<$9.73 & & 0.0 & $<$0.76 & & 0.0 & $<$0.17 & \\
J2055+2539   & 35.3 & 16.06 $\pm$ 10.90 & 1.23 $\pm$ 0.76 & 23.3 & 1.53 $\pm$ 0.42 & 4.89 $\pm$ 0.75 & 0.0 & $<$0.13 & \\
J2124$-$3358 & 16.0 & 16.75 $\pm$ 12.17 & 1.83 $\pm$ 0.70 & 56.6 & 2.41 $\pm$ 0.54 & 2.34 $\pm$ 0.37 & 0.0 & $<$0.21 & \\
J2229+6114   & 4.2 & $<$49.68 & & 0.0 & $<$1.36 & & 0.0 & $<$0.28 & \\
J2238+59     & 2.5 & $<$55.91 & & 0.0 & $<$1.54 & & 0.0 & $<$0.38 & \\
\enddata

\tablenotetext{\textit{a}}{The spectral parameters of the Crab Nebula are derived using \citet{crab}.}
\tablenotetext{\textit{b}}{The spectral parameters are derived assuming a uniform disk morphology as described in \citet{velax}.}

\tablecomments{Results of the maximum likelihood spectral fits for the off-pulse emission of LAT gamma-ray pulsars (see Section~\ref{ana}). 
The off-pulse spectra were fit with a power-law model (photon index $\Gamma$ and photon flux $F$) 
assuming a point-source at the position of the pulsar. The results for the fits in the three energy bands are reported. 
The test statistic ($TS$) for the source significance is provided in Columns 2 (0.1 -- 1 GeV)), 5 (1 -- 10 GeV) and 8 (10 -- 100 GeV). 
The photon flux $F$ for each energy band is reported in Columns 3, 6, and 9; it is replaced by a 2$\sigma$ upper limit when $TS <$ 25 (assuming a photon index $\Gamma$=2). 
Columns 4, 7 and 10 list the photon index $\Gamma$ for each energy band when $TS \ge$ 25. Only 
statistical uncertainties are reported on the photon flux and the photon index.}

\end{deluxetable}
\end{landscape}

\begin{landscape}
\begin{deluxetable}{lcrrcr}
\tablecaption{Spectral fitting of pulsar wind nebula candidates with low energy component.
\label{tab:spec}}
\tablewidth{0pt}

\tablehead{
\colhead{PSR} & \colhead{$G_{0.1-100}$} & \colhead{$\Gamma$} & \colhead{$E_{\rm cutoff}$} & \colhead{$TS_{\rm cutoff}$}\\
 & \colhead{($\rm 10^{-12} \,erg \ cm^{-2}\, s^{-1}$)} & & \colhead{(GeV)} & &
}
\startdata
J0034$-$0534 & 7.33 $\pm$ 2.01 $\pm$ 1.30 & 0.62 $\pm$ 1.05 $\pm$ 0.27 & 0.7 $\pm$ 0.48 $\pm$ 0.10 & 9.0 \\
J0633+1746 & 544.01 $\pm$ 13.91 $\pm$ 54.58 & 1.51 $\pm$ 0.06 $\pm$ 0.12 & 1.41 $\pm$ 0.14 $\pm$ 0.09 & 247.2 \\
J1813$-$1246 & 116.24 $\pm$ 22.92 $\pm$ 79.28 & 2.65 $\pm$ 0.14 $\pm$ 0.26 &  & 1.2 \\
J1836+5925 & 349.64 $\pm$ 16.04 $\pm$ 28.05 & 1.33 $\pm$ 0.10 $\pm$ 0.06 & 1.60 $\pm$ 0.25 $\pm$ 0.04 & 99.8 \\
J2021+4026 & 737.14 $\pm$ 21.77 $\pm$ 125.06 & 1.87 $\pm$ 0.06 $\pm$ 0.20 & 2.24 $\pm$ 0.37 $\pm$ 0.51 & 110.2 \\
J2055+2539 & 12.23 $\pm$ 6.14 $\pm$ 6.09 & 0.30 $\pm$ 1.40 $\pm$ 0.69 & 0.43 $\pm$ 0.31 $\pm$ 0.07 & 22.4 \\
J2124$-$3358 & 13.27 $\pm$ 3.02 $\pm$ 2.77 & 0.88 $\pm$ 0.74 $\pm$ 0.34 & 1.71 $\pm$ 1.06 $\pm$ 0.59 & 10.4 \\

\enddata

\tablecomments{Results of the maximum likelihood spectral fits for pulsars showing a significant signal in their off-pulse 
at low energy. The fits used an exponentially cutoff power-law model 
with the energy flux $G_{0.1-100}$, photon index $\Gamma$ and cutoff energy $E_{\rm cutoff}$ given in columns 2, 3 and 4. 
The first errors represent the statistical error on the fit parameters, while the second ones are the systematic uncertainties 
as discussed in section~\ref{analow}. The significance of an exponential cutoff (as compared to a simple power-law) is indicated 
by $TS_{\rm cutoff}$ in column 5. A value $TS_{\rm cutoff}$ $< 9$ indicates that the two models are comparable and we report the fit parameters assuming a simple power-law model. }
\end{deluxetable}
\end{landscape}


\end{document}